\input epsf	

 
\magnification= \magstep1  
\tolerance=1600 

\parskip=5pt 
\baselineskip= 6 true mm  \mathsurround=1.4pt
\font\smallrm=cmr8
\font\smallit=cmti8

\def\a{\alpha}
\def\b{\beta}
\def\g{\gamma} 
\def\d{\delta} \def\D{\Delta}
\def\e{\varepsilon}

\def\k{\kappa}
\def\l{\lambda} \def\L{\Lambda} 
\def\m{\mu}
\def\f{\phi} \def\F{\Phi}
\def\n{\nu}
\def\j{\psi} 
\def\r{\varrho}
\def\s{\sigma} 
\def\t{\tau}
\def\th{\theta}  \def\Th{\Theta}

\def\w{\omega} \def\W{\Omega}

\def\bra{\langle} \def\ket{\rangle}
\def\dd{{\rm d}}
\def\cl{\centerline} \def\ni{\noindent}
\def\pa{\partial}\def\ra{\rightarrow}
\def\tl{\tilde}
\def\scrunch{\baselineskip=10 pt \smallrm}
\def\toe{\hfil\break\vskip-18pt}

\def\secbreak{\vskip12pt plus .7in \penalty-200\vskip 0pt plus-.5in} 
\def\prefbreak#1{\vskip#1\penalty-50\vskip-#1}
 
\newcount\noteno
\def\fn#1#2{\footnote{$^#1$} {\scrunch #2 \toe}}
\def\fnd#1{\fn{\dagger}{#1}}
\def\fndd#1{\footnote{$^\ddagger$} {\scrunch #1 \toe}}

\def\numfn#1{\global\advance\noteno by 1 
        \footnote{$^{\the\noteno}$} {\scrunch #1 \toe}}

\def\ddef{\ {\buildrel {\rm def}\over =}\ }

\def\fract#1#2{{\textstyle{#1\over#2}}}
\def\half{\fract12}
\def\qu{\ {\buildrel {\displaystyle ?} \over =}\ }
\def\bal{$\bullet$} 
\def\ref#1{$^{\hbox{\smallrm #1}}$}
 
\def\ex#1{e^{\textstyle{#1}}}
\def\HH{{\cal H}} \def\OO{{\cal O}} \def\LL{{\cal L}}
\def\NN{{\cal N}} \def\TT{{\cal T}} \def\DD{{\cal D}}
\def\low#1{{\vphantom{\big)}}_{#1}} \def\llow#1{{\vphantom{\big)}}_{\!#1}}
\def\In{{\rm in}}    \def\Out{{\rm out}}
 \def\Inout{{\rm in\atop out}} 
 \def\g44{\Big(1-{2M\over r}\Big)}
 
\def\lap{\setbox0=\hbox{$<$}\,\raise .25em\copy0\kern-\wd0\lower.25em\hbox{$\sim$}\,}
\def\glt{\setbox0=\hbox{$>$}\,\raise .25em\copy0\kern-\wd0\lower.25em\hbox{$<$}\,}
\def\gap{\setbox0=\hbox{$>$}\,\raise .25em\copy0\kern-\wd0\lower.25em\hbox{$\sim$}\,}

\def\tds{{\tl\pa}^{\,2}}
\def\in{\low{\rm in}}    \def\out{\low{\rm out}} \def\BH{\low{\rm BH}}
\def\tot{^{\rm tot}}
\def\Gbar{\raise.13em\hbox{--}\kern-.35em G}

\vglue 1truecm
\rightline{THU-96/26}
\rightline{gr-qc/9607022}
 
\vfil {\bf\cl{ THE SCATTERING MATRIX APPROACH}\cl{ FOR THE QUANTUM BLACK
HOLE}}\bigskip \cl{\it an overview}
\vfil

\cl{G. 't Hooft } \bigskip \cl{Institute for Theoretical Physics}
\cl{Utrecht University , P.O.Box 80 006} \cl{3508 TA Utrecht, the
Netherlands} \vfil 
{\bf Abstract} \medskip{\narrower{
If one assumes the validity of conventional quantum field theory in
the vicinity of the horizon of a black hole, one does not find a
quantum mechanical description of the entire black hole that even
remotely resembles that of conventional forms of matter; in contrast
with matter made out of ordinary particles one finds that, even if
embedded in a finite volume, a black hole would be predicted to have a
strictly continuous spectrum.
 
Dissatisfied with such a result, which indeed hinges on assumptions
concerning the horizon that may well be wrong, various investigators
have now tried to formulate alternative approaches to the problem of
``quantizing" the black hole. We here review the approach based on the
assumption of quantum mechanical purity and unitarity as a starting
point, as has been advocated by the present author for some time,
concentrating on the physics of the states that should live on a black
hole horizon. The approach is shown to be powerful in not only
producing promising models for the quantum black hole, but also new
insights concerning the dynamics of physical degrees of freedom in
ordinary flat space-time.}\smallskip}

\vfil\eject                     

\ni{\bf 1.  INTRODUCTION}\medskip
More and more physicists working on attempts to reconcile the theory of
general relativity with the postulates of quantum mechanics are
becoming aware of the obstinate problems arising when gravitational
collapse might occur in microscopic systems. In classical general
relativity there is nothing wrong with the ``black hole solution",
whose properties can be precisely calculated using ordinary laws of
physics, and which may arise naturally in astronomical objects with
masses several times that of our Sun.  Indeed it is not hard to
indicate the classical initial conditions that inevitably will lead to
a black hole, as we will briefly summarise in Sections 2 and 3. In
particular one observes that the emergence of a horizon has nothing to
do with small scale physics, and therefore one generally expects that
black hole properties, at scales large compared to the Planck length,
are indeed represented accurately by these classical calculations\ref{1, 2}.

In the classical case, the emergence of a singularity at the origin of
space-time is in no way problematic. The region where it occurs if
physically unobservable and so, for practical purposes, in particular when
one wants to predict ``experimantal observations", this singularity is of no
importance. Throughout this paper it will receive little attention (see
however section~17).

One would have expected that also the quantum mechanical properties of a
black hole should follow naturally by applying large scale physics. Only the
space-time region at the side of the observer, the ``physical side" of the
horizon, should be relevant. Indeed one can calculate accurately the
quantummechanical effects near a large black hole, as seen by an outside
observer, by first studying what an infalling observer would experience, and
then performing the appropriate general coordinate transformation. As is to
be expected from quantum mechanical calculations, one finds ``probabilities":
chances that particles of certain types, with certain momenta, energies or
other quantum numbers, emerge at certain places.  It is when one wants to
interpret these outcomes in terms of some Schr\"odinger equation for the
black holes as a whole, that the first genuine problems emerge\ref{3, 4}.

Does a black hole, enclosed in a {\it  finite\/} volume of space,
possess a discrete or a continuous spectrum? We will argue in Section~7
that one would expect the spectrum to be discrete, but more careful
analysis of Hawking's result suggests a continuum to be inevitable,
even if interactions are taken into account\ref{5, 6}. This situation is
fundamentally different from what we have in any ordinary form of
matter. It would imply, among other things, that a black hole cannot
disappear at the end of its evaporation process; a {\it  remnant\/}
should then stay behind (see for instance Ref\ref{ 7}).  Still equipped
with a continuous spectrum, such remnants would defy any description in
terms of conventional physical laws. No spin-statistics theorem would
apply to them, and no thermodynamics can be worked out for such
objects.

Whether remnants are acceptable or not in a logically coherent theory can now
be disputed. Many physicists argue that there is as yet no conflict. Black
holes will always behave in an exceptional way. Others however maintain that
in the presence of remnants that can be abundantly created virtually, the
construction of consistent theories that fully predict the physical effects
at the Planck length will be impossible, a situation that is unprecedented in
any other branch of physics.  More to the point is the observation that there
are many reasons to question the validity of the arguments that led to these
results; more likely quantum gravity will require drastically different
techniques, and the conclusion that black holes terminate their lives as
remnants appears to be highly premature.

As will be argued in much more detail in this review article, on the
one hand we have observations on large black holes, made by distant
observers; these can probably be deduced entirely\ref{2, 6} by applying
conventional physics (though even here there could be surprises\ref8).
On the other hand, however, questions concerning their spectrum will
require new ultra short distance physics for their answers.

What makes this problem so important is that its resolution, whichever
it will be, will strongly affect our views on all other laws of physics
at the Planck scale as well.  A theory with virtual black hole remnants
will look quite different from a world where conventional quantum
mechanics holds unabatedly at the tiniest distance scales. But also if
conventional quantum mechanics, free from decoherence effects from
black holes, can be rescued, the world at the Planck scale will be
different from conventional physical theories as we will further
discuss (Section~16). Yet, it is this latter avenue that this author
strongly prefers for further investigation. It leads to rich physics,
connected in several curious ways to string theories. It may thus
provide for an improved interpretation of string theory.  In short, our
assumption is that Schr\"odinger equations can be used universally for
all dynamics in the universe, including the Planck scale\ref{9, 10}.  This
assumption can obviously not be {\it  proved} from any more basic first
principle; it is a first principle itself. Principles of this sort are
indeed badly needed for the construction of a logically coherent theory
for Planck scale physics.  It can of course also be wrong. We have
nothing further to say about that.

The technique that is applied is simply {\it assuming\/}\ref{10} that
black holes form a discrete spectrum just as any other physical system;
the density of states is then approximately prescribed by the entropy
law. Combining this assumption to known laws of physics, numerous
properties of the $S$-matrix elements involving black holes can
actually be derived. The procedure is completely straightforward and
free from ambiguities, at least in principle.  Its elaboration is the
subject of this article.

Relatively little attention will be paid to the generalizations of the
Schwarz\-schild solution: the Reissner-Nordstrom and Kerr-Newman
solutions\ref{11, 12}.  These are the solutions obtained when electric charge and
angular momentum are added to the hole. Conceptually, generalization of
everything we say to these cases should be straightforward. Recent
literature pays a considerable amount of attention to the {\it
extremal} solutions\ref{13}, which are the ones obtained when electric charge,
angular momentum, or a combination of both, take the maximally allowed
values.  As models these cases are quite interesting, but in the
present paper they are given less priority. Our motivation for this is
that these limiting cases cannot be representative for the general
black hole; it could well be that these solutions will be unattainable
in ``realistic" experiments at the Planck scale. In view of this, they
might show rather pathological behavior without upsetting the logic of
the underlying theory (more about this at the end of Section~2).
Conclusions based on a study of these solutions alone might therefore
be misleading.

For the same reason we will not pay much attention to dilaton black
holes.  Dilaton fields might be important at the Planck scale, but
appear not to exist at macroscopic scales; macroscopic black holes
should be logically consistent without dilatons. There are several
recent developments which are not yet in a stage that they can be
thoroughly reviewed here. One is the attempts to count black hole
states in the context of string theory\ref{14}.  It is claimed that the
results strongly support our views, since the numbers obtained agree
with the value of the Bekenstein entropy\ref{15}.  Their interpretation
in terms of modes living on the horizon is still obscure however, and
we have no idea how to look against general coordinate transformations
for these states, which is the problem addressed in our paper.

An interesting suggestion was made by Carlip\ref{16}, who claims that
states can be counted for a black hole in 2+1 dimensions. His result
sounds remarkable since the theory he considers is a topological one.
If the characteristics of the states counted by him can be further
identified, this would be a valuable piece of information for our
theories. The philosophy adhered to in our paper is to start our
analysis from the other end: we insist on first asking the question
what the physical effects are that we wish to describe. Only after
these have been identified, one may decide which mathematical tricks
should be used to speed up and simplify calculations.

The paper is organised as follows.  The introductory parts,
Sections~1--10, contain only rather elementary material, but are needed
as a background for the remainder (most of it can be skipped by a more
informed reader). The coordinate frames used are exhibited in
Section~2. Then, in Section~3, a picture is sketched of a black hole
formed by matter. This picture should make it clear that conventional
physical laws can never be such that black holes could be avoided --
they are legitimate solutions of classical, ({\it i.e.},
non-quantummechanical) physics. For some special conventions in our
notation, see Section~4.

The standard Hawking-Unruh effect, and its relation to Bogolyubov
transformations of creation and annihilation operators, are explained
in Sections~5 and~6. In Section~7, the standard argument is given 
explaining why we expect only a finite density of physically allowed
states near a horizon, by deriving the entropy of a black hole. From
the ``brick wall model", in Section~8, we learn that ordinary quantum
fields already provide the required number of states if we limit these
to exist only outside a ``brick wall", at distance $h$ from the
horizon.

All this has given us a detailed picture of the {\it statistical}
behavior of particles near a black hole, but the laws of physics at a
microscopic scale that should be held responsible for these effects are
still obscure. In Section~9 we explain that, in order to zoom in to
these laws, we need to understand the gravitational back reaction of
the Hawking radiation process. The first step is the Aichelburg-Sexl
metric (Section~10), which is then exploited in Section~11, where we
make a first attempt to set up a description of the Hilbert space of
black hole horizon states. Non-gravitational forces must be
accommodated for as well (Sections~12 and~13), but the results fail to
account for the finiteness of density of states on a given area of the
horizon. What is needed for this is the incorporation of the {\it
sideways} gravitational forces. Their source is the transverse momentum
operator (Section~14), and attempts are made to incorporate these effects
in the horizon operator algebra (Section~15).

The most urgent motivation for this work is its possible relevance to
modelling physics at the Planck scale in ordinary flat space-time, as
is further elaborated on in Section~16. In Section~17 we briefly speculate 
on how to proceed from here.

\secbreak

{\ni\bf 2. THE SCHWARZSCHILD METRIC}\medskip 
The space-time metric of a stationary, non-rotating and electrically neutral
black hole is the {\it  Schwarzschild metric}: 
$$\dd s^2\,=\,-\Big(1-{2M\over r}\Big)\dd t^2+{\dd r^2\over
1-2M/r}+r^2\dd\W^2\,,\eqno(2.1)$$ where
$$\dd\W^2\,\equiv\,\dd\th^2+\sin^{\!2}\th\,\dd\f^2\,.\eqno(2.2)$$
Furthermore, $M$ stands for $Gm\BH $, where $G$ is Newton's constant and
$m\BH $ is the black hole mass. Often we will employ the {\it  Kruskal\/
coordinates}\ref{12}, which we will write as $(x,\,y,\,\th,\,\f)$, with
$$\eqalign{\Big({r\over 2M}-1\Big)e^{r/2M}\,&=\,xy\,;\cr
e^{t/2M}\,&=\,x/y\,.\cr}\eqno(2.3)$$
In terms of these coordinates we have (see Fig.~1)
$$\eqalign{{\dd x\over x}+{\dd y\over y}\,&=\,{\dd r\over
2M\big(1-2M/r\big)}\,;\cr {\dd x\over x}-{\dd y\over y}\,&=\, {\dd t\over
2M}\,;\cr \dd s^2\,&=\,{32M^3\over r}e^{-r/2M}\,\dd x\dd y + r^2\dd
\W^2\,.\cr}\eqno(2.4)$$
The apparent singularity at the horizon, $r=2M$, has disappeared. The only true
singularities are at the curves $xy=-1$, where $r=0$. The region $\{ x>0,\
y>0\}$ is the ``outside region", the only region from which distant observers
can obtain any information. The line $y=0$, where $r=2M$, is the ``future
horizon"; the line $x=0$ where also $r=2M$, is the ``past horizon".

\midinsert\epsffile{thfig1.ps}\scrunch\cl{Figure  1. Varous coordinates used to
describe the Schwarzschild metric.}
\cl{The local light cones are oriented everywhere as indicated.} \endinsert 
In the region $r\approx2M$ one can write the metric as $$\dd
s^2\,\approx\,{16M^2\over e}\dd x\dd y+4M^2\dd\W^2\,\eqno(2.5)$$ and with the
coordinate substitution $$\eqalign{{4M\over\sqrt e}\,x\,&=\, Z +T  \,,\cr
{4M\over\sqrt e}\, y\,&=\, Z -T  \,,\cr 2M\th\,&=\,\half\pi+X \,,\cr
2M\f\,&=\,Y \,,\cr}\eqno(2.6)$$ 
close to the origin, one finds that in terms of these coordinates 
space-time  is approximately flat: 
$$\dd s^2\,\approx\,-\dd T ^2+\dd Z ^2+\dd X ^2+\dd Y ^2\,.\eqno(2.7)$$ 
The transformation 
$$Z \, =\,\r\cosh \t \,,\qquad T \, =\,\r\sinh \t \,, \eqno(2.8)$$
brings us back to the Schwarzschild coordinates (close to the horizon), 
apart from normalization factors: 
$$t/2M\,=\,2\t \,,\qquad 8M(r-2M)\,=\,\r^2\,.\eqno(2.9)$$ 
The description of a flat space-time (2.7) in terms of the coordinates
(2.8) is called ``Rindler space"\ref{17}. We see that close to the
horizon, the Schwarzschild coordinates $r$ and $t$ behave as Rindler
space coordinates.

The derivation of more general black hole solutions carrying electric
charge and/or angular momentum can be found in the
literature.\ref{11, 12} One finds
$$\dd s^2\,=\,-{\D\over Y}(\dd t+a\sin^2\th\dd\f)^2+{\sin^2\th\over
Y}(a\dd t+ (r^2+a^2)\dd\f)^2+{Y\over\D}\dd
r^2+Y\dd\th^2\,,\eqno(2.10)$$  with
$$\D\,=\,r^2-2Mr+Q^2+a^2\ ,\qquad Y\,=\,r^2+a^2\cos^2\th\,,\eqno(2.11)$$
where $J=Ma$ is the angular momentum, and $Q$ is the electric charge in
conveniently chosen units. The two horizons, 
$r_+$ and $r_-$, are the solutions of the
equation $\D(r)=0$. 

{\it Extreme black holes\/} are represented by the case
$$Q^2+a^2\,=\,M^2\,,\eqno(2.12)$$ in which case $r_+-r_-\downarrow0$.
They (together with versions including dilaton fields) are presently
being studied in string theory scenarios. Two aspects of the extreme
black holes should be kept in mind:\par\ni One: both horizons are
separated from the observer by an infinite distance, since $r_+=r_-$
implies that $\int_{r_+}^{r_1}\dd r\sqrt{Y/\D}$ diverges
logarithmically.  \par\ni Two: the two horizons may appear to coincide
in the extreme limit, but in reality they do not.  The (timelike)
distance between $r_+$ and $r_-$ is in the extreme limit~(2.12):

$$\lim_{\e\ra 0}\int_{r_-}^{r_+}\dd r\sqrt{M^2+a^2\cos^2\th\over\e^2-
(r-M)^2}\,=\,\pi\sqrt{M^2+a^2\cos^2\th}\,,\eqno(2.13)$$
which remains macroscopic at large $M$. 
Thus, for an observer who stays close to the horizon there is no extreme limit;
for him, the horizon at $r=r_+$ continues to look like an ordinary horizon,
which, even in the extreme limit, stays far away from the horizon at $r=r_-$.
This is why the author believes
that macroscopic extreme black holes are  physically ill-defined limits; it would
be a mistake to treat an extreme horizon as just one horizon. Most likely,
this limit can only be understood if one first manages to control the regular case,
not vice versa.
The string theory results reported in ref\ref{ 14}
are very encourageing, but, as stated earlier, they are not yet in a stage that
they can be reviewed here.
  \secbreak

{\ni\bf 3. BLACK HOLE FORMED BY MATTER}\ref{18, 19}\medskip 
So far the stationary solution was described. To see that black holes can
actually be formed by ordinary matter we have to study time-dependent
solutions. A class of such solutions can easily be found. Take the case that
we have spherical symmetry, and matter that only consists of particles moving
radially inwards and outwards with the speed of light, without interacting
(here, the outward moving particles are included for later considerations; in
this section they will beignored from Eq.~(3.9) onwards).  The metric is
taken to be 
$$\dd s^2\,=\,2A(x,y)\dd x\dd y + r^2(x,y)\dd\W^2\,,\eqno(3.1) $$ 
where $A$ and $r$ are functions of $x$ and $y$ yet to be determined. Matter
only contributes to the components $T_{xx}$ and $T_{yy}$ of the
energy-momentum tensor, whereas all other components, including the trace
$T_\m^\m$, are zero:
$$T_{xy}\,=\,T_{\th\th}\,=\,0\,,\eqno(3.2)$$ implying for the Ricci tensor
$R_{\m\n}$: $$R_{xy}\,=\,R_{\th\th}\,=\,0\,.\eqno(3.3)$$  If we introduce a new
function $M\,=\,M(x,y)$ and write $$A\,=\,{2r\,r_xr_y\over r-2M}\,,\eqno(3.4)$$
where $r_x$ stands for ${\pa r\over\pa x}\Big|_y$, etc., then the equations
(3.3) become remarkably simple: $$r_{xy}\,=\,{2M\,r_xr_y\over r(r-2M)}\,,\qquad
M_{xy}\,=\,{-2M_xM_y\over r-2M}\,.\eqno(3.5)$$ They both can be integrated once
to give $$2M_xr_x\,=\,g(x)\Big(1-{2M\over r}\Big)\,,\qquad
2M_yr_y\,=\,h(y)\Big( 1-{2M\over r}\Big)\,,\eqno(3.6) $$ where $g(x)$ and
$h(y)$ are arbitrary functions. Since $$8\pi G T_{xx}\,=\,{2g(x)\over
r^2}\,\quad{\rm and}\quad 8\pi G T_{yy}\,=\,{2h(y)\over r^2}\,,\eqno(3.7)$$ we
must demand that $g(x)\ge 0$ and $h(y)\ge 0$. These functions are determined by
fixing the flux of ingoing and outgoing particles at infinity. At $r\ra\infty$
we write $$\eqalign{r\,\ra\,{1\over\sqrt2}(x+y)\,&;\qquad
t\,=\,{1\over\sqrt2}(x-y)\,,\cr \dd s^2\,&\ra\,\dd r^2-\dd
t^2+r^2\dd\W^2\,.\cr}\eqno(3.8)$$ The ingoing matter depends only on the $x$
coordinate and the outgoing matter on the $y$ coordinate (see Fig.~2a).

\midinsert\epsffile{thfig2.ps}\narrower\scrunch\ni Figure 2. a) Spherically
symmetric configuration of matter radially moving inward and outward with the
speed of light.  b) Spherically symmetric black hole formed by radially
inmoving lightlike matter.\endinsert

Now take the case $h(y)=0$, which corresponds to the case that there is no
outgoing matter.  Since we do not want  $r_y$ to vanish, Eq.~(3.6) implies
$$M(x,y)\,=\,M(x)\,,\eqno(3.9)$$
and since at infinity $r_x\,\ra\,{1\over\sqrt2}$ we get $$g(x)\,=\,\sqrt2\,
{\dd M(x)\over\dd x}\,.\eqno(3.10)$$ Note that at $t\ra-\infty$ we have
$$\eqalign{T_{tt}\,&=\,T_{rr}\,=\,T_{rt}\,=\,T_{tr}\,=\,\half\ T_{xx}\,,\qquad
T_{xy}\,=\,T_{yy}\,=0\,,\cr E\,&=\,4\pi\int r^2T_{tt}\dd r\Big|_t\ =\ \int
{1\over 2G}g(x)\dd r\Big|_t\ \ra\ \int{1\over\sqrt2}{\pa M\over\pa x}\dd
r\Big|_t\cr &=\,\int\dd x{\dd M\over G\dd x}\,=\,{M\over G}\,.\cr}\eqno(3.11)$$
 We see from this solution that matter imploding in a spherically symmetric
way, and with the speed of light, without any further interactions, always
collapses into a black hole. The horizon appears as soon as $r=2M$, at which
point the local situation is still completely regular, so that there is no
classical mechanism to be found that could block this collapse from
happening. The clearest situation is the case where we have a thin
spherically symmetric shell of imploding matter:
$$ M(x)\, =\,M_0\Th(x-x_0)\ ,\qquad g(x)\,=\,\sqrt2\,M_0\d(x-x_0)\,,\eqno(3.12)$$ 
where we choose $x_0>0$. At $x<x_0$ we have flat space-time: 
$$ r\,=\,{1\over\sqrt2}(x+y)\ ,\qquad A=1\qquad (x<x_0)\,.\eqno(3.13)$$ 
At $x>x_0$ we define a new $y$ variable, $\tl y $, by $$r(x_0,y)\,=\,{1\over
\sqrt2}(x_0+y)\,;\qquad \tl y\,=\,{1\over x_0}\Big(
{r(x_0,y) \over 2M}-1\Big)\ex{r(x_0,y)/2M}\,.\eqno(3.14)$$ 
Then at $x>x_0$ we recover 
$$x\tl y \,=\,\Big({r(x,\tl y )\over2M}-1\Big)\ex{r(x,\tl y )/2M}\ ;\qquad
M(x,\tl y )\,= \,M_0\,.\eqno(3.15)$$ 
Thus we have flat space-time inside the shell and the Schwarzschild solution
outside. See Fig.~2b. The horizon is the sheet $\tl y =0$, corresponding to
$\,r(x_0,y)=2M\,$ or $\,y=y_1=2\sqrt2\,M-x_0\,$. The horizon opens up for the
first time at the point $S$ defined by $r(x,y_1)=0\,$ or
$$x\low S\,=\,-2\sqrt2\,M+x_0\ ,\qquad y\low S\,=\,2\sqrt2\,M-x_0\,,\eqno(3.16)$$
but this situation becomes more complicated as soon as deviations from 
spherical symmetry are introduced.

The freely falling thin shell of matter can also be treated when this
matter moves slower than the speed of light. Again one gets a flat
space-time inside, and the Schwarzschild solution outside.  The horizon
starts at a point $S$ well inside  the flat region, and the shell moves
through this horizon in a region where everything is still completely
regular.  Thus one concludes that if the equation of state describes
non-interacting matter and the initial condition is spherically
symmetric an implosion into a black hole will often be
inevitable\fnd{The arguments that black hole formation is also
inevitable in more generic circumstances without spherical symmetry are
more delicate and will not further be discussed here.}.  In principle,
a sufficiently large and dilute dust cloud can obey these conditions.
The importance of such considerations  is that this way one can
convince oneself that black holes must exist as part of the spectrum of
physically realizable states. The question ``what is the smallest
possible black hole?" must be asked and answered in any viable theory
of quantum gravity.\fndd{There is evidence that even in {\smallit classical
physics}, black holes that are arbitrarily small {\smallit can\/} be made,
provided the initial data are sufficiently carefully chosen, see
ref\ref{ 20}.} It is quite plausible that at the Planck scale no clear
distinction will be possible between black holes and ordinary
particles, but if this is the case we must ask how the ``unification"
between black holes and ordinary matter can be realized in mathematical
terms.  \secbreak

{\ni\bf 4. CONVENTIONS AND NOTATION}\medskip 
In the equations of the previous two sections we have 
$$M\,\equiv\,Gm\BH \,,$$ where $m\BH $ is the mass of the 
black hole and $G$ the gravitational constant. Later we will not choose 
units such that $G$ is normalized to one, because this would leave 
factors $16\pi$ in the Lagrangian. In analogy with the notation
$\hbar=h/2\pi$, one could propose the notation
$$\Gbar\,=\,8\pi G\,.$$ In any case, we will pick units such that 
$$G\,=\,1/8\pi\,,\eqno(4.1)$$ 
in view of equations (10.13) and (10.23).
  
In previous publications, various choices have been made for factors $\sqrt 2$
in the lightcone coordinates. Omitting such factors does no harm in general
considerations of particular details, but leads to ambiguities in the
definitions when later comparisons are made.  For this reason we here put these
factors in, so that as a consequence some numerical factors ar not quite the
same as in previous work. We are not yet in a situation that comparison with
the phenomenology of the real world can be made so that as yet factors of 2 are
of no direct importance, but we will try to be precise.

The coordinates 
$$(t,\,r,\,\th,\,\f)\ \leftrightarrow\ (x,\,y,\,\th,\,\f)\ \leftrightarrow\
(T,\,X,\,Y,\,Z)$$ are as in Sect. 2. Here $(T,\,X,\,Y,\,Z)$ are coordinates in
terms of which a small region near the horizon has an approximately Minkowskian
metric (2.7). In addition, we will use $$\eqalign{x^+\,=\,{1\over\sqrt2}(Z+T)\
=\,{\r\over\sqrt2}\ex\t\ ;&\qquad x^-\,=\,{1\over\sqrt2}(Z-T)\
=\,{\r\over\sqrt2}\ex{-\t}\,,\cr &\tl x\,=\,\pmatrix{X\cr
Y\cr}\,,\cr}\eqno(4.2)$$ and 
$$\eqalign{k^+\,=\,{1\over\sqrt2}(k^3+k^0)\ ;&\qquad
k^-\,=\,{1\over\sqrt2}(k^3-k^0)\ ;\cr \tl k\,=\,\pmatrix{k^1\cr
k^2\cr}\,,&\qquad k^0\,=\,+\sqrt{\tl k^2+m^2+{k_3}^2}\,,\cr}\eqno(4.3)$$
so that 
$$(k\cdot X)\,=\,k^+x^-+k^-x^++\tl k\cdot\tl x\ ,\quad{\rm and}\quad \dd
s^2\,=\,2\dd x^+\dd x^-\,+\,{\dd\tl x}^2\ ,\eqno(4.4)$$ and the mass shell
condition reads
$$k^-\,=\,-(\tl k^2+m^2)/2k^+\,.\eqno(4.5)$$

The {\it  physical region} of a classical, permanent black hole is defined by
$$\eqalign{\r\,>\,0\qquad{\rm or}\qquad r\,>\,2M &\qquad\rm or \cr  x\,>0\ ;\quad
y\,>\,0&\qquad\rm or \cr Z\,>\,|T|&\qquad\rm
or \cr x^+\,>\,0\ ;\quad x^-\,>\,0&\,. \cr}\eqno(4.6)$$
In Sections~11--14, heavy use is made of operators that we call the
``momentum density", of various sorts. These are naturally related to
the energy-momentum tensor in light cone coordinates, but they are not
quite the same thing.  The relations are as follows:
$$\eqalignno{P_\In(\tl x)\,&=\,\int T_{++}(\tl x,x^+,x^-)\dd x^+\big|_{
x^-=0}\,;&(4.7)\cr
P_\Out(\tl x)\,&=\,-\int T_{--}(\tl x,x^+,x^-)\dd x^-\big|_{x^+=0}\,;&(4.8)\cr
\tl P^a_\In(\tl x)\,&=\,\int T_{a+}(\tl x,x^+,0)\dd x^+\,;&(4.9)\cr
\tl P^a_\Out(\tl x)\,&=\,-\int T_{a-}(\tl x,0,x^-)\dd x^-\,.&(4.10)\cr}$$
Here, the transverse index $a$ takes the values 1 or 2. The minus sign
has its origin in our definition of the lightcone minus components.
\secbreak

\ni{\bf 5. THE RINDLER SPACE TRANSFORMATION}\ref{6, 21}\medskip
Consider the Minkowski coordinate frame $\{T,X,Y,Z\}$, or $\{T, {\bf X}\}$ for short,
and a scalar field $\F(T,{\bf X})$. Let this field simply obey a Klein-Gordon
equation, $$(\pa^2-m^2)\F\,=\,0\,.\eqno(5.1)$$ 
The quantum theory is written in the Heisenberg representation, which means
that the states $|\j\ket$ are space-time independent, but the fields are
operators depending both on space and on time. Usually, a complete set of
solutions of (5.1) is written in terms of the Fourier modes with respect to the
Minkowski space coordinates, and one gets
$$\eqalignno{\F({\bf X},T)\,&=\,\int{\dd^3{\bf k}\over\sqrt{2k^0({\bf
k})(2\pi)^3}}\Big(a({\bf k})e^{i{\bf k\cdot X}-ik^0T}+a^\dagger({\bf
k})e^{-i{\bf k\cdot X}+ik^0T}\Big)\,,&(5.2)\cr \dot\F({\bf
X},T)\,&=\,\int{-ik^0\dd^3{\bf k}\over\sqrt{2k^0({\bf k})(2\pi)^3}}\Big(a({\bf
k})e^{i{\bf k\cdot X}-ik^0T}-a^\dagger({\bf k})e^{-i{\bf k\cdot
X}+ik^0T}\Big)\,.&(5.3)\cr}$$ 
Here $k^0({\bf k})=\sqrt{{\bf k}^2+m^2}$, and the
transformation from $a$ and $a^\dagger$ to $\F$ and $\dot\F$ has been designed
such that the following commutation rules are maintained: $$\eqalignno{[\F({\bf
X},T),\,\F({\bf X}',T)]\,=\,0\,,\qquad&[\F({\bf X},T),\,\dot\F({\bf
X}',T)]\,=\,i\d^3({\bf X}-{\bf X}')\,,&(5.4)\cr [a({\bf k}),\,a({\bf
k}')]\,=\,0\,,\qquad&[a({\bf k}),\,a^\dagger({\bf k}')]\,=\,\d^3({\bf k}-{\bf
k}')\,.&(5.5)\cr}$$
Not only do these commutation rules ensure that 
$a^\dagger$ and $a$ act as creation and annihilation operators, but also the
time dependence in (5.2) and (5.3) implies that the objects created and
annihilated carry an amount of energy equal to $k^0$.

The operator $H_M$ that generates  boosts in the time coordinate $T$, 
$${\pa\F\over\pa T}\,=\,-i[\F,H_M]\,,\eqno(5.6)$$
is the Minkowski-Hamiltonian
$$H_M\,=\,\int\HH({\bf X})\dd^3{\bf X}\,=\,
\int\dd^3{\bf k}\,k^0({\bf k})a^\dagger({\bf k})a({\bf k})\,,\eqno(5.7)$$

We need first the transition to light-cone coordinates (4.2), and we define 
$$a({\bf k})\sqrt{k^0}\,=\,a_1(\tl k,k^+)\sqrt{k^+}\,.\eqno(5.8)$$
Since 
$${\pa k^+\over\pa k^3}\Big|_{\tl
k}\,=\,{1\over\sqrt2}\Big(1+{k^3\over\sqrt{\tl
k^2+{k_3}^2+m^2}}\Big)\,=\,{k^+\over k^0}\,,\eqno(5.9)$$
the new operators $a_1$, $a_1^\dagger$ are normalized by 
$$[a_1(\tl k, k^+),\, a_1^\dagger(\tl k',{k^+}')]\,=\,\d^2(\tl k-\tl k')\d(k^+ 
-{k^+}')\,.\eqno(5.10)$$ 
Thus one can write $$\F(T,{\bf X})\,=\,A(x^+,x^-,\tl x)+A^\dagger(x^+,x^-,\tl x)\,
,\eqno(5.11)$$ with 
$$A(x^+,x^-,\tl x)\,=\,\int\dd^2\tl k\int_0^\infty{\dd k^+\over\sqrt{2k^+(2\pi)^3}} 
a_1(\tl k,k^+)e^{i(\tl k\cdot\tl x+k^+x^-+k^-x^+)}\,,\eqno(5.12)$$

In terms of the Rindler space coordinates $\{\t,\r,\tl x\}$ of Eq.~(2.8) the 
Klein Gordon equation (5.1) reads
$$\big[(\r\pa_\r)^2-\pa_\t^{\ 2}+\r^2(\tl\pa^2-m^2)\big]\F\,=\,0\,.\eqno(5.13)$$
Solutions periodic in $\t$ are\fndd{The notations used here differ slightly from 
our notation in Ref\ref{28}.}:
$$\eqalign{\F_\w\,=&\,K(\w,\,\half\m\r e^\t,\,\half\m\r e^{-\t})\,e^{i\tl k\cdot\tl x}\,=\cr 
=&\,K(\w,\,\half\m\r,\,\half\m\r)\,e^{i\tl k\cdot\tl x-i\w\t}\,,\cr}\eqno(5.14)$$ 
where $\m^2=\tl k^2+m^2\,$ and 
$$K(\w,\a,\b)\,=\,\int_0^\infty{\dd s\over s}s^{i\w}e^{-is\a+i\b/s}\,.\eqno(5.15)$$
This function $K$ obeys 
$$K(\w,\,\s\a,\,\b/\s)\,=\,\s^{-i\w}K(\w,\a,\b)\,,\eqno(5.16)$$ 
and can be expressed in terms of familiar Bessel and Hankel functions.
Eq.~(5.15) is readily obtained by taking in 4-dimensional space-time one of
the plane wave solutions: $k^+=-k^-=\m/\sqrt2\,$, which are rewritten in terms of the
coordinates $\r,\,\t$ of Eq.~(4.2), and then 
Fourier transformed with respect to $\t$. It is not difficult to verify directly (using 
partial integration in $s$) that the partial differential equation (5.13) is 
obeyed.

We will now normalize the Fourier components of the fields $\F$ with respect to
the $\t$ coordinate as follows:
$$A(x^+,x^-,\tl x)\,=\,\int_{-\infty}^\infty\dd\w\int{\dd^2\tl k\over\sqrt{2(2\pi)^4}}K
(\w,\,\half\m\r,\,\half\m\r)e^{i\tl k\tl x-i\w\t}a_2(\tl k,\w)\,,\eqno(5.17)$$ 
since then the new operators $a_2$ are identified as
$$a_2(\tl k,\w)\,=\,(2\pi)^{-1/2}\int_0^\infty{\dd k^+\over\sqrt{k^+}}
a_1(\tl k,k^+)\ex{i\w\ln\big({k^+\sqrt2\over\m}\big)}\,,\eqno(5.18)$$
a Fourier transformation whose inverse is:
$$a_1(\tl k,k^+)\sqrt{k^+}\,=\,(2\pi)^{-1/2}\int_{-\infty}^\infty\dd\w 
a_2(\tl k,\w)\ex{-i\w\ln\big({k^+\sqrt2\over\m}\big)}\,.\eqno(5.19)$$
Substituting (5.19) into (5.12) gives us back (5.17) and (5.15).  
With the normalization factors chosen in (5.17)--(5.19), the operators $a_2$
and $a_2^\dagger$ again obey
$$[a_2(\tl k,\w),\,a_2^\dagger(\tl k',\w')]\,=\,\d^2(\tl k-\tl k')\d(\w-\w')\,, 
\eqno(5.20)$$ 
and therefore one might think that they could serve as annihilation and 
creation operators in Rindler space. But then one has to realize that in 
the integrals (5.17), $\w$ can take negative values. The corresponding 
operators $a(\w)$ would annihilate a {\it  negative} amount of Rindler energy.
To cure this situation we first need to know some simple properties of the 
function $K$:

First of all one has:
$$K^*(\w,\a,\b)\,=\,K(-\w,-\a,-\b)\,.\eqno(5.21)$$
Let now $\a>0$ and $\b>0$. In the definition Eq.~(5.15) the integrand is
bounded in the region Im$(s)\le 0$. Therefore one can rotate the integration
contour as follows:
$$s\,\ra\,s\,e^{-i\f}\ ,\qquad 0\le\f\le\pi\,.\eqno(5.22)$$
Taking the case $\f=\pi$ we find
$$K(-\w,\a,\b)\,=\,\int_0^\infty{\dd s\over s}s^{i\w}
e^{-\pi\w}e^{is\a-i\b/s}\,=\, e^{-\pi\w}K^*(\w,\a,\b)\qquad{\rm 
if}\quad\a>0,\ \b>0\,,\eqno(5.23)$$ and similarly one has:
$$K(-\w,\a,\b)\,=\, e^{+\pi\w}K^*(\w,\a,\b)\qquad\quad {\rm if}\quad 
\a<0,\ \b<0\,.\eqno(5.24)$$
This allows us to rearrange the integral over $\w$ in Eq.~(5.17) to obtain 
for $\F=A+A^\dagger$ in the region $\r>0$:\prefbreak{.5in}
$$\F\,=\,\int_0^\infty\dd\w\,e^{-i\w\t}\int{\dd\tl k\,e^{i\tl k\tl x}\over 
\sqrt{2(2\pi)^4}}K(\w, \half\m\r,\half\m\r)\Big(a_2(\tl k,\w)+e^{-\pi\w}
a_2^\dagger(-\tl k,-\w)\Big)\ +\,{\rm h.c.}\eqno(5.25)$$
In the opposite quadrant of Rindler space, where $\r<0$, we have:
$$\F\,=\,\int_0^\infty\dd\w\,e^{-i\w\t}\int{\dd\tl k\,e^{i\tl k\tl x}\over 
\sqrt{2(2\pi)^4}}K(\w, \half\m\r,\half\m\r)\Big(a_2(\tl k,\w)+e^{+\pi\w}
a_2^\dagger(-\tl k,-\w)\Big)\ +\,{\rm h.c.}\eqno(5.26)$$

At this point it is opportune to define the new creation and annihilation 
operators $a_I,\,a_{II},\,a_I^\dagger,$ and $a_{II}^\dagger$ applying the 
following Bogolyubov transformation, when $\w>0$:
$$\pmatrix{a_I(\tl k,\w)\cr a_{II}(\tl k,\w)\cr a_I^\dagger(-\tl k,\w)
\cr a_{II}^\dagger(-\tl k,\w)\cr}\,=\,{1\over 
\sqrt{1-e^{-2\pi\w}}}\pmatrix{1&0&0&e^{-\pi\w}\cr 0&1&e^{-\pi\w}&0\cr 
0&e^{-\pi\w}&1&0\cr e^{-\pi\w}&0&0&1\cr}\pmatrix{a_2(\tl k,\w)\cr a_2(
\tl k,-\w)\cr a_2^\dagger(-\tl k,\w)\cr a_2^\dagger(-\tl k,-\w)\cr}\,.
\eqno(5.27)$$
This way, at $\r>0$ the field $\F$ only depends on $a_I$ and $a_I^\dagger$, 
and at $\r<0$ only on $a_{II}$ and $a_{II}^\dagger$, whereas the 
normalization has been chosen again such that 
$$\eqalign{[a_I(\tl k,\w),a_I^\dagger(\tl k',\w')]\,&=\,[a_{II}(\tl k,\w),
a_{II}^\dagger(\tl k',\w')]\,=\,\d(\w-\w')\d^2(\tl k,\tl k')\,;\cr
[a_I,\,a_{II}]\,&=\,[a_I,\,a_{II}^\dagger]\,=\,0\,.}\eqno(5.28)$$

This Bogolyubov transformation does {\it  not\/} directly affect the Rindler
Hamiltonian, the latter being the generator of a boost in the $\t$ direction:
$$\eqalign{{\pa\F\over\pa\t}\,&=\,-i[\F,\,H_R]\,,\cr
-i\w a_2(\tl k,\w)\,&=\,-i[a_2(\tl k,\w),\,H_R]\,,\cr}\eqno(5.29)$$ where $$
\eqalign{H_R\,&=\,\int_{-\infty}^\infty\dd\w\,\w a_2^\dagger(\tl k,\w)a_2
(\tl k,\w)\cr&=\,\int_0^\infty\dd\w\,\w\Big(a_I^\dagger(\tl k,\w)a_I(\tl k,\w)\,-\,
a^\dagger_{II}(\tl k,\w)a_{II}(\tl k,\w)\Big)\cr
&=\,H_R^I\,-\,H_R^{II}\,.\cr}\eqno(5.30)$$
Thus we observe that Hilbert space is separable into two factor spaces: 
$\HH=\HH_I\otimes\HH_{II}$. The space $\HH_I$ is described by the Hamiltonian
$H_R^I$ and $\HH_{II}$ is described by the Hamiltonian $-H_R^{II}$. One may
also verify that, apart from a possible additive contribution of the vacuum,
if $$H_M\,=\,\int\dd^3{\bf X}\, \HH_M\Big(\F(0,{\bf X}),\vec\pa\F(0,{\bf X}),
\dot\F(0,{\bf X})\Big)\,,\eqno(5.31)$$ then $$H_R^I\,=\,\int_{\r>0}\dd^3{\bf
X}\,\r\HH_M\ ;\qquad H_R^{II}\,=\,\int_{\r<0} \dd^3{\bf
X}\,|\r|\HH_M\,.\eqno(5.32)$$ Consequently, all observables made of fields in
quadrant $II$ commute with $H_R^I$ and {\it  vice versa}.

The Rindler- or Boulware vacuum state $|0,0\ket$ is defined by
$$a_I|0,0\ket\,=\,a_{II}|0,0\ket\,=\,0\eqno(5.33)$$ This is not the same as
the vacuum experienced by a freely falling observer, who is said to experience
the Minkowski- or Hawking vacuum $|\W\ket$, which obeys $$A(T,{\bf
X})|\W\ket\,=\,a_2(\tl k,\w)|\W\ket\,=\,0\,.\eqno(5.34)$$
It is not difficult to express this state in terms of the basis generated by 
$a_I$ and $a_{II}$:
$$\eqalign{a_I(\tl k,\w)|\W\ket\,&=\,e^{-\pi\w}a_{II}^\dagger(-\tl k,\w)
|\W\ket\,,\cr a_{II}(\tl k,\w)|\W\ket\,&=\,e^{-\pi\w}a_I^\dagger(-\tl k,\w) 
|\W\ket\,,\cr}\eqno(5.35)$$ so that 
$$|\W\ket\,=\,\prod_{\tl k,\w}\sqrt{1-e^{-2\pi\w}}\sum_{n=0}^\infty|n\ket\low{I}
|n\ket\low{II}e^{-\pi n\w}\,,\eqno(5.36)$$ where the square root is added for 
normalization.

Notice that $$H_R|\W\ket\,=\,(H_R^I-H_R^{II})|\W\ket\,=\,0\,,\eqno(5.37)$$ 
which confirms that $|\W\ket$ is Lorentz invariant; remember that $H_R$ 
is the generator of Lorentz boosts. 

If one does not have the means to observe any features at $\r<0$ this implies
that one only has at one's disposal operators $\OO$ composed of the fields in
region $I$, that is, the operators $a_I$ and $a_I^\dagger$. These act only in
the factor space $\HH_I$ but proportional to the identity operator in
$\HH_{II}$:
$$\OO\Big(|\j\ket_I|\j'\ket_{II}\Big)\,=\,|\j'\ket_{II}\,\Big(\OO|\j
\ket_I\Big)\,.\eqno(5.38)$$ 
Let us limit ourselves momentarily to a single point $(\tl k,\w)$.  There the
expectation value for such an operator in the state $|\W\ket$ is
$$\eqalign{\bra\W|\OO|\W\ket\,&=\,(1-e^{-2\pi\w})\sum_{n_1,n_2}\low{II}
\bra n_1| \,\low{I}\bra n_1|\OO|n_2\ket\low{I}\,|n_2\ket\low{II}
\ex{-\pi\w(n_1+n_2)}\cr &=\,\sum_{n\ge0}\low I\bra n|\OO|n\ket\low I e^{-2\pi
n\w}(1-e^{-2\pi\w})\,,\cr &=\,{\rm
Tr}\big(\OO\,\r\low\W\big)\,,\cr}\eqno(5.39)$$
where $\r\low\W$ is the {\it  density matrix} $C e^{-\b H_I}$
corresponding to a thermal state at the temperature\ref6
$T=1/2\pi$. Note that in Rindler space time, energy and temperature 
are dimensionless. If we scale with the appropriate factor $4M$ as 
in Eq.~(2.9) we find the {\it  Hawking temperature}, 
$$T\low H\,=\,1/8\pi M\,=\,1/8\pi Gm\BH \,.\eqno(5.40)$$

It should be emphasized here that the above is a rather straightforward
calculation of a result that can also be obtained in a number of other,
different ways. Several of these alternative derivations are more
elegant and direct, for instance one may observe that the factors $e^{\pm
i\pi\w}$ arise from a rotation over $\pi$ in Euclidean space, and that the
thermal nature of the result is linked to the fact that the Euclidean extension
of the Schwarzschild solution is periodic in $it$ with period $8\pi M$. The
derivation given in this section was chosen because of the directness of the
{\it  physical\/} arguments, which are independent of the Euclidean properties of
the solutions.
\secbreak

\ni{\bf 6. HAWKING RADIATION}\medskip
S. Hawking\ref{3} was the first to arrive at an apparently inevitable
conclusion, drawn by starting from the results sketched in Sect. 5:
that a black hole will emit black body radiation, precisely
corresponding to the temperature (5.40). One immediately concludes that
there cannot be a lower bound to the black hole mass much higher than
the Planck mass, since, once it is placed in a high enough vacuum, the
black hole will loose energy:  $${\dd M\over\dd
t}\,=\,C(T)T^4R^2\,\approx\,-{C'\over M^2}\ ;\qquad M(t)\,\approx\,
C''(t_0-t)^{1/3}\,.\eqno(6.1)$$ Here, $C$, $C'$, etc., are all constants of
order one in natural units. Furthermore, if $S$-matrix theory would
apply to them, the black hole instability would shift the pole $s_j$
generated by black hole of type $j$, in the $S$-matrix, into the complex plane:
$$s_j\,\approx\,{m_{j,}^2}\BH-i C_j M\low{\rm Pl}^2\,,\eqno(6.2)$$
where $C_j$ is of order one in natural units. 
 
Numerous authors investigated this chain of arguments and found that
they generally agree. Indeed, suspicions that there could be deviations
from this result, involving non-thermal fluctuations, or even a
different temperature\ref{8} are most often not taken very seriously.
Caution however is called for.  We must underline that here we are
dealing with a purely theoretical prediction which, whether we like it
or not, is based on assumptions that cannot all be verified directly,
plausible as they may seem.  Black holes emitting quantum radiation
have never been observed experimentally, and indeed it is conceivable
that either Quantum Mechanics or General Relativity, or both, might
break down precisely at the horizon, regardless how large the horizon
is. Such fears are not totally unfounded, as we will argue later in
this paper, but it is very illuminating first to derive what the
consequences of the conventional theories would be.  Let us therefore
begin with rederiving Hawking's result as precisely as we can, stating
what assumptions were made.

Consider the black hole just formed out of imploding matter. The model pictured
in Fig.~2b is quite suitable for our considerations, and it will not be
difficult to convince oneself that the details of the imploding matter are
totally irrelevant for the arguments. As in the previous section, we consider
quantized fields in this space-time in a Heisenberg picture (which means that
the states $|\j\ket$ are time-independent and that the operators referring to
the various observers are defined as a function of their location $x$ in this
space-time). Take an operator-valued field $\F(x)$ interacting no other way than
gravitationally (later we can relax also this condition). If we are interested
in observations of particles leaving the black hole at late time $t$, this means
that we are interested in modes that propagate along lines such as the line $L$
in Fig.~2b. These modes are rapidly oscillating as functions of the coordinate
$y$ but practically constant as functions of $x$. Now consider these modes as
they cross a Cauchy surface near the point labled $S$ in Fig.~2b. At this point
we assume that the energetic modes that we are interested in are not populated
by any particle, or, the associated annihilation operators $a({\bf k})$ are
assumed to yield zero when acting on the vacuum (for short: ``annihilate the
vacuum"). Because the field $\F(x)$ obeys field equations, we can also consider
some later Cauchy surface at time $t$, near the letter $L$ in Fig.~2b. We can
now calculate how the operators $a({\bf k})$ act as experienced by an observer
there. To do this, we may safely assume that also the observer moving in along
the curve $A$ in Fig.~2b sees no super-energetic particles along the outgoing
curve $L$. Thus we can ignore any of the effects of the imploding material that
created the black hole itself, and concentrate on the metric of an eternal
black hole, Fig.~1, where the same curves $A$ and $L$ may be considered (see
Fig.~1).

Since the wave fronts that we wish to study are tightly compressed, it is
permitted to concentrate on a tiny region of the horizon, where the Minkowski
coordinates $T,X,Y,Z)$ are appropriate. The Hamiltonian $H_M$ that serves as
the generator of boosts in the time variable $T$ can be written as
$$H_M\,=\,\int\HH({\bf X})\dd^3{\bf X}\, .\eqno(6.3)$$  

Now for the late observer watching the radiation going out along curve $L$ the
relevant coordinates are $\r$ for space and $\t$ for time, where $\r$ and $\t$
are defined as in Eq.~(4.2).  Since all our considerations are now in the upper
part of space-time we can use the exact Schwarzschild metric (2.1) or its
Kruskal notation (2.4). A boost in the new time coordinate $\t$ is generated by
the {\it  Rindler\/} Hamiltonian $H_R$, defined as in Eq.~(5.32).  This, apart
from the factor $4M$, is the time coordinate used by the late observer.  What
the results of the previous section imply, is that if the observer near the
point $S$ observes few if any ultra-energetic particles, we can say that the
state we are dealing with is the Hawking vacuum $|\W\ket$. The field operators
$\F$ in the matter-free region outside the horizon still act exactly the same
way on this state as they do in the Rindler space treatment of Sect. 5. If we
now perform the general coordinate transformation to the Rindler variables, and
from there to the familiar Schwarzschild coordinates $t,r,\th,\f$, we find
that the appropriate field variables there consist of the creation and
annihilation operators $a_I$ and $a_I^\dagger$. The state $|\W\ket$ becomes the
density matrix state $\r\low\W$ derived from Eq.~(5.39).

This argument is highly independent of the way the black hole was
formed. In case the collapse took place in a less symmetric way, or at
various steps and intervals, one still finds that an observer falling
in the black hole should observe the Hawking vacuum state there, and
this necessarily leads to the density matrix $\r\low\W$. In particular,
one could assume that the collapsing matter was in a {\it pure} quantum
state, and even in that case, the outgoing radiation appears to be
mixed according to the matrix $\r\low\W$. The question
to be asked is how literally this result is to be taken. One could
conclude \item{$i$)} that black holes must be fundamentally different
from other objects in nature. They do not obey a single Schr\"odinger
equation (which after all would allow pure states to evolve only into pure
states), but in stead obey probabilistic equations of motion that are
not purely quantum mechanical\ref{4, 22}.  \par\ni According to this view,
a more basic theory at the Planck scale would show no quantum
mechanical features of the familiar kind. Scenarios where this {\it
does\/} happen have been proposed, but the present author is skeptical
towards such proposals. The reason is not that they would be logically
impossible (although difficulties with energy conservation were
claimed\ref{23}), but rather that unitarity violation would be practically
impossible to build in a comprehensive theory of Planckian evolution
such a way that Quantum Mechanics at the atomic scale would result
naturally. Existing scenarios show a complete lack of determinism (even
in the sense of a Schr\"odinger equation) at the Planck scale and hence
do not allow us even in principle to make any sound physical prediction
for quantities such as the fine structure constant.  Instead, perhaps,
\item{$ii$)} black holes do obey a Schr\"odinger equation, but this
equation requires knowledge of all inaccessible observables behind the
horizon, so that a black hole forms an infinitely degenerate state. In
this case the black hole can never decay completely\ref7, but it decays into
stable, infinitely degenrate, final states with masses of the order of
the Planck mass, called {\it remnants}. Alternatively, however, one may
suspect that \item{$iii$)} The density matrix derivation depended on
certain hidden assumptions of a statistical nature, such that the
answer may be correct in a statistical sense, but more precise
treatments may yield a purely quantum mechanical description of a black
hole that nevertheless has only a finite degeneracy. This is the
scattering matrix assumption, which we will further investigate from
section~11 onwards.\par\ni
\ni Thus, one expects the system as a whole to react just as any other
physical system does: when it absorbs infalling material, or even just
infalling radiation, it should react some way or other, and enter into
a state that is orthogonal to what it would have evolved into if the
infalling material had been in a different mode or totally absent. This
is just the experimentally observed fact that all known evolution laws
in small-scale physics can be written in terms of a {\it  unitary\/}
evolution matrix. It is hard to understand how the world at the scale
of ordinary atomic and elementary particle physics could behave quantum
mechanically and evolve in a unitary way, if quantum mechanics were not
at the basis of the laws of dynamics at the smallest distance scales.

In appears that the derivation of the density matrix $\r\low\W$ in
Eq.~(5.39) cannot be exactly right, since it inplies that infalling
material of whatever variety should not affect the outgoing radiation
at all (linearized quantum field theory was used). This would violate
unitarity.

There are (at least) three different reasons to expect that the density
matrix $\r\low\W$ has to be replaced by a pure state. First, as already
explained, the {\it  result\/} of the computations that provided the
density matrix $\r\low\W$ is difficult to accept at all levels of
rigor.

Secondly, there is the back reaction: of course outgoing radiation
{\it  does\/} depend on ingoing material. After all, the black hole
mass and angular momentum, and hence also the surrounding space-time
metric, are affected by whatever went in, and therefore also the
location of the horizon, the temperature and other observable
features.  However, taking into account back reaction, is {\it  not\/}
sufficient to resolve the unitarity problem. Even if we keep track of
{\it everything\/} that went in, the outgoing radiation nevertheless
appears to be quantum mechanically mixed. Our point, however, is that
the Rindler space picture may be more fundamentally flawed: the density
matrix that was derived requires states in the original Minkowski space
that feature particles with tremendous energies (with respect to the
Kruskal coordinates), both ingoing and outgoing. These particles should
interact extremely strongly gravitationally; it would be incorrect to
ignore this super strong interaction, as we  will see in section~10.

Thirdly, we claim that the derivation depends on assumptions that need
not be true. In Eq.~(5.39) it was tacitly assumed that the expectation
value of an operator $\OO$ should be calculated by averaging over all
unseen modes $|n_2\ket$. In statistical physics, of course, this is the
standard procedure, so much so that its applicability is nearly never
questioned. But is it the right way to treat information that
disappeared across a gravitational horizon? If we take the laws of
General Relativity literally the answer is {\it yes}. In an accelerated
frame, we also cannot see what went through the acceleration horizon,
and there the rule would definitely be to average over all possible
unseen states. Yet it could be that General Relativity in this respect
has to be modified near a horizon. One proposal was made by this
author\ref8 in 1984: it could be that the double Hilbert space
$\HH_I\otimes\HH_{II}$, mentioned following Eq.~(5.30), itself has to
be regarded as a space of density matrix elements. This might look like
a more economical and perhaps more natural rule whenever a
gravitational field produces a horizon. The resulting theory differs
from the standard one by a factor two in the temperature $T\low H$.

A theory such as General Relativity could be interpreted as essentially
consisting of the following ingredients: {\it different} physical
situations are compared (for instance a region of space-time with a
gravitational field and an other region without gravitational field).
Then a {\it transformation law} is formulated enabling us to calculate
what happens in one system (the one with the gravitational field), if
we know how to calculate what happens in the other region (the one
without gravitational field). In the conventional theory, the
transformation involved here is nothing but a coordinate
transformation. Quantum mechanical Hilbert space is assumed not to be
affected. Now we claim that in the presence of a horizon the
transformation rule could be more complicated. For instance, a {\it
state} $|\j\ket$ in flat space could be transformed into a {\it density
matrix\/} $\r$ near the horizon (this then would give a factor two
difference in $T\low H$)

Unfortunately, this particular proposal does not resolve the unitarity
problem, and implementing it in a more comprehensive theory for black
holes turned out to be even harder than for the conventional view. We
were unable however to rule it out completely, and all we can do at
present is to take this heretic theory as a warning sign that not all
may seem to be as one might expect. In principle, the transformation law
could be even more intricate than the simple theories just mentioned.
In this paper, we will leave this question open. At some places, we
speculate on Hawking's original formalism, which is by far the most
likely to be correct.  It could be, however, that the transformation
from Minkowski space to Rindler space is not a one-to-one mapping in
terms of states in Hilbert space, because of the divergence due to
infinitely energetic particles.

\secbreak

\ni{\bf 7. BLACK HOLE THERMODYNAMICS}\medskip
The fact that the radiation emitted, as described by Eq.~(5.40), is {\it
thermal}, opens up the possibility to approach this phenomenon from a
thermodynamical point of view. Taking $m\BH $ to be the energy and $T=T\low H$ the
temperature, one readily derives the {\it entropy}~$S$: 
$$T\dd S\,=\,\dd m\BH \,;\qquad\dd S\,=\,8\pi G m\BH \dd
m\BH \,;\qquad S\,=\,4\pi G m^2\!\BH +C\,,\eqno(7.1)$$
where $C$ is an unknown integration constant, to be referred to as the
``entropy normalization constant".\fn*{There may well be subdominant $m\BH $-
dependent terms needed in Eq.~(7.1), which would imply that $C$ is not truly
constant.} 

For the general Kerr-Newman solution (see sect.~2 and\ref{11, 12}), where
$J=Ma$ is the angular momentum, $Q$ the electric charge in convenient
units, $r_+$ the radius of the horizon, $\F$ the electric potental on
the horizon, and $\W$ the angular velocity at the horizon, these
expressions generalize into:
$$\eqalignno{r_\pm=M\pm\sqrt{M^2-a^2-Q^2}\,;&\qquad\f={Qr_+\over
r_+^2+a^2}\,; \qquad \W={a\over r_+^2+a^2}\,;&(7.2a)\cr T\low
H&\,=\,{r_+-r_-\over 4\pi(r_+^2+a^2)}\,;&(7.2b)\cr G\,T\dd S \,=\,\dd
M-\W\dd J-\f\dd Q\,;&\qquad G\,S
\,=\pi(r_+^2+a^2)\,+\,G\,C\,.&(7.3a,b)\cr}$$ It is indeed not difficult
to check that Eq.~(7.3b) solves the differential equation (7.3a). The
expressions (7.2) for the Kerr-Newman solution will not further be
derived here.

However, deriving the value of the entropy normalization constant $C$  from
conventional physical arguments is a great challenge, and it has not yet been
accomplished. In fact, not only is $C$ still unknown, but we also have no idea
where in the range between, say, $10^{-40}$ and $10^{+40}$, it might be;
indeed, one may even dispute whether it is finite and well-defined at all. One
could also suspect that $C$ depends on properties of the black hole (other
than $Q$ or $a$) that are determined by its history ($Q$ and $a$ dependence
are excluded by Eq.~(7.3a)), but we consider such speculations as unlikely.

It is important to note that the expression obtained for the entropy $S$,
apart from the integration constant, is always equal to $\fract14  A/G$, where
$A$ is the {\it area} of the horizon, a finding that will be very much at the
center of our discussions.  For dilaton black holes, which will not be
discussed here, the situation may be more complicated.

We would like to compare the black hole with an inflatable balloon
filled with gas molecules. Through a small opening, molecules may enter
or leave. When it is large, and contains many molecules, the
statistical, thermodynamic treatment is appropriate.  But when it is
small, with only one or two molecules left, it should be more accurate
to solve the Schr\"odinger equation. {\it Formally}, a treatment in
terms of pure quantum states should always be possible. The black hole
should only differ from the toy balloon by the fact that its quantum
states seem to be labled by spinlike variables, separated by
approximately the Planck length, on the horizon.

Let us return to the pure Schwarzschild black hole. Connecting the entropy to
the {\it density of quantum mechanical states}\ref{15}, must be done with
considerable care, since there will be two kinds of divergences: at the
horizon and at spacelike infinity. In fact, one may very well question the
mere {\it existence\/} of such quantum levels. This, however, is the key
assumption of this paper: not only is the quantum mechanics of black holes
meaningful, it can also be {\it derived}, and the constant $C$ in Eqs.(7.1)
and (7.3b) is finite and of order one (apart from subdominant terms).  In
order to enable us to judge the relation between the entropy just derived, and
the density of quantum states, we now present a direct argument concerning the
density of states, an argument that will also show that any infinities at the
{\it horizon\/} must be absorbed in $C$, but the ``infrared" infinities arising
from spacelike infinity should be excluded; the latter represent the radiation
field far from the black hole.

The spectral density of a black hole can be derived from its Hawking
temperature by applying time reversal invariance\ref9. We have to our disposal both
the {\it emission rate\/} (the Hawking radiation intensity), and the {\it
capture probability\/}, or the effective cross section of the black hole for
infalling matter.

The cross section $\s$ is approximately:
$$\s\,=\,2\pi r_+^2\,=\,8\pi M^2\,,\eqno(7.4)$$
and slightly more for objects moving in slowly. The emission probability $W\dd
t$ for a given particle type, in a given quantum state, in a large volume
$V=L^3$ is:
$$W\dd t\,=\,{\s({\bf k}) v\over V}\,\ex{-E/T}\dd t\,,\eqno(7.5)$$
where $\bf k$ is the wave number characterizing the quantum state, $v$ is the
particle velocity, and $E$ is its momentum.

Now we {\it assume\/} that the process is also governed by a Schr\"odinger
equation. This means that there exist quantum mechanical transition amplitudes,
$$\eqalign{\TT\in\,&=\,\BH\!\bra M+GE|\ \ |M\ket\BH|E\ket\in \,,\cr{\rm and}\qquad 
\TT\out\,&=\,\BH\!\bra M|\,\out\!\bra E|\ \ |M+GE\ket\BH\,,}\eqno(7.6)$$
where the states $|M\ket\BH$ represent black hole states with mass $M/G$, and
the other states are energy eigenstates of particles in the volume $V$. In terms
of these amplitudes, using the so-called Fermi Golden Rule, the cross section 
and the emission probabilities can be written as
$$\eqalignno{\s\,&=\,|\TT\in|^2\r(M+GE)/v\,,&(7.7)\cr
W\,&=\,|\TT\out|^2\r(M){1\over V}\,.&(7.8)\cr}$$
where $\r(M)$ stands for the level density of a black hole with mass $M$. The
factor $v^{-1}$ in Eq.~(7.7) is a kinematical factor, and the factor $V^{-1}$
in $W$ arises from the normalization of the wave function.

Now, time reversal invariance relates $\TT\in$ to $\TT\out$. To be precise, all
one needs is $PCT$ invariance, since the parity transformation $P$ and charge 
conjugation $C$ have no effect on our calculation of $\s$.
Dividing the  expressions (7.7) and (7.8), and using (7.5), one finds:
$${\r(M+GE)\over\r(M)}\,=\,\ex{E/T}\,=\,\ex{8\pi ME}\,.\eqno(7.9)$$
This is easy to integrate:
$$\r(M)\,=\,\ex{4\pi M^2/G + C}\,=\,\ex S\,.\eqno(7.10)$$

Eq.~(7.10) may be rewritten as
$$\r(M)\,=\,2^{A/A_0}\,,\eqno(7.11)$$
where $A$ is the horizon area and $A_0$ is a fundamental unit of area,
$$A_0\,=\,4G\ln 2\,.\eqno(7.12)$$
This suggests a spin-like degree of freedom on all surface elements of size $A_0$.

Extension to the more general Kerr-Newman solutions is straightforward; 
one has to take into account that the Hawking emission intensity $\ex{-E/T_H}$
is modified by chemical potential terms when the particles considered carry 
an electrical charge $q$ or  angular momentum $j$ (in conveniently chosen units):
$$W\,\propto\,\ex{(-E+\W j+\f Q)/T}\,.\eqno(7.13)$$
One again finds Eq.~(7.3a), that is to be integrated to obtain (7.3b), where $S$
is the logarithm of the level density.

As stated earlier, the importance of this derivation is the fact that the
expressions used as starting points are the {\it actual} Hawking emission rate
and the {\it actual} black hole absorption cross section. This implies that,
if in more detailed considerations divergences are found near the horizon,
these divergences should not be used as arguments to adjust the relation
between entropy and level density by large renormalization factors.
Furthermore, the Golden Rule argument can be used only to deal with one
emitted particle at the time. Hence, we should not take the outside volume $V$
so large that the dominant emission mode contains very many particles.
Therefore, any divergences found when the outside volume is taken to infinity
should be subtracted.

\secbreak
\ni{\bf 8. THE BRICK WALL MODEL}\ref9\medskip
Attempts to compute the entropy normalization constant $C$ from field
theory near black hole horizons are doomed to failure\fn*{One important
exception is the result obtained by Carlip\ref{16}, who derived
the black hole entropy in 2+1 dimensions, in a theory with a
cosmological constant. There are two reasons why this is an exceptional
case: one is that in the limit where the cosmological constant is small
his black hole is either very large or very heavy. His
universe can only contain one black hole at most; second: there are no
other particles in his model, the degrees of freedom are purely
topological and gravitational. Nevertheless, the result is of great
interest.} (unless we would be content
with the value infinity). This statement we derive from the following
argument\ref{24}. A quantum field theory may easily exhibit global additive
conservation laws such as baryon number conservation (or, more realistically,
conservation of $B-L$, where $L$ is lepton number, but for simplicity, we
will refer to the conserved quantity as baryon number). Such conservation
laws are connected to global $U(1)$ symmetries {\it via} Noether's theorem,
in which case there is no long range gauge field coupled to this conserved
current.  Any quantum mechanical amplitudes derived within the framework of
such a theory should display the same global symmetries. If $C$ is finite,
however, our amplitudes $\TT$ of Eq.~(7.6) cannot possibly share the symmetry.
Since the cross section $\s$, Eq.~(7.4), for baryon capture is fixed, it
should be possible for the black hole to capture unlimited amounts of
baryons, whereas the number of baryons and antibaryons emitted in the Hawking
process should always average out to be equal, since there will be no
chemical potential term (as in Eq.~(7.13)) favoring one above the other.
Thus, on the one hand, one can increase the baryon number of a black hole
indefinitely, without increasing its mass. If, on the other hand, one has
only a finite number of quantum states below a given mass value, one has a
conflict unless baryon number is ill-defined for these states, which means
that the symmetry is broken. Such a symmetry breaking can only come from
physical mechanisms that were not part of the given quantum field theory.

We would now like to see explicitly what goes wrong if one attempts to do the
calculation anyway. The answer is that quantum field theory does not give to
the black hole a discrete spectrum, but gives it a continuous spectrum of
states, {\it i.e.}, $C=\infty$. Since we clearly have a discrete spectrum of
particles in the $GeV$ range, this result would imply that the continuum
should begin somewhere, presumably at a mass value comparable to the Planck
mass. This ``lightest black hole" would not be able to decay, since this
would require a transition from a continuous spectrum into a discrete
spectrum, which in general is impossible quantummechanically. In short, one
would predict the existence of so-called {\it black hole remnants}.

Remnants, which would form an infinite set of different species, all at about
the same mass value, would as such disobey conventional rules of quantum
statistics, and thermodynamics would not apply to them. We will not be able
to {\it prove} that this is unacceptable, but, in our opinion, it is unlikely
that such behavior can be accommodated in an airtight quantum theory.  More
to the point, however, is the remark that Quantum Field Theory in a black
hole background is not more than an approximation method, in which
gravitational back reaction was neglected. In a four-dimensional theory, this
approximation may well be not good enough to conclude anything at all about
the spectrum.

In this Section, we now present a model in which only {\it low energy} quantum
fluctuations of the fields are taken into account. We apply quantum field
theory op to some point $r_1$ close to the horizon: $r_1=r_++h,\ h>0$.  For
simplicity we only consider scalar fields $\F_i(r,\th,\f,t)$, whose only interaction
is the gravitational one with the metric; generalization
towards spinor, vector or even perturbative gravitational field excitations
will be straightforward. To simplify things, we just represent all those by
giving the fields $\F_i$ a multiplicity $N$, so $i=1\dots,N$. At $r=r_1$ we
impose a boundary condition:
$$\F_i(r,\th,\f,t)=0\qquad{\rm if}\qquad r\le r_1\,.\eqno(8.1)$$
The quanta of the fields will be given a temperature $T=T\low H$. The 
question one may ask is: which value should one assign to the cutoff
parameter $h$, such that the entropy of this system precisely takes the
value (7.1), so that the density of quantum states corresponds to (7.10)?
We will need an infrared cutoff in the form of a box with radius $L$:
$$\F_i(r,\th,\f,t)=0\qquad{\rm if}\qquad r\ge L\,.\eqno(8.2)$$

To determine the thermodynamic properties of this system, we first
compute the energy levels $E(n,\ell,\ell_3)$ of the bosons $\F_i$.
The Lagrange density $\LL$ in the metric (2.1) is
$$\LL({\bf x},t)\,=\,\g44^{-1}\pa_t\F_i^2\,-\,\g44\pa_r\F_i^2\,-\,r^{-2}\pa_\W
\F_i^2\,.\eqno(8.3)$$
The total Lagrangian is $\int_{r_1}^L\dd r\int\dd\W r^2
\LL(r,\W,t)$, 
so that the field equation for modes with angular momentum quantum numbers 
$\ell$, $\ell_3$ and energy $E$ reads:
$$\g44^{-1}E^2\F+{1\over r^2}\pa_r r(r-2M)\pa_r\F-\Big({\ell(\ell+1)\over r^2}
+m^2\Big)\F\,=\,0\,.\eqno(8.4)$$
To enable us here to apply a WKB approximation, we first must smoothen 
the singularity of the second term in (8.4) at $r=2M$, since this singularity 
is too close to the cutoff point $r-r_1$. Therefore, temporarily write
$$r-2M\,=\, e^\s \,,\eqno(8.5)$$
so that
$$\bigg[rE^2+{1\over r^2}\pa_\s r\pa_\s- e^\s \Big({\ell(\ell+1)\over r^2}+
m^2\Big)\bigg]\F\,=\,0 \qquad(r=2M+ e^\s )\,.\eqno(8.6)$$
In terms of the $\s$ coordinate we see that the oscillatory behavior of
$\F$ is well approximated by
$$\ex{\pm i\int \k(\s)\dd\s}\,=\,\ex{\pm i\int k(r)\dd r}\,,\eqno(8.7)$$
where $\k(\s)$ is determined by
$$rE^2-{1\over r}\k(\s)^2- e^\s \Big({\ell(\ell+1)\over r^2}+m^2\Big)\,=\,0\,,
\eqno(8.8)$$
so that\ref9
$$k(r)^2\,=\,e^{-2\s} \k(\s)^2\,=\,\g44^{-2}E^2-\g44^{-1}\Big({\ell(\ell+1)
\over r^2}+m^2\Big)\,,\eqno(8.9)$$
as long as the right hand side of this equation is positive; let us take
$k(r)=0$ as soon as the right hand side is negative.

The energy spectrum of Eq.~(8.6) is now given by
$$\pi n\,=\,\int_{r_1}^L \dd r\, k(r,\ell,E)\,,\eqno( 8.10)$$
where the quantum numbers $n>0$, $\ell$ and $\ell_3=-\ell,\dots,\,\ell$ 
are  integers. The total number $\n$ of wave solutions with energy not
exceeding $E$ is then given by 
$$\eqalign{\pi\n\,&=\,\int(2\ell+1)\dd\ell\, \pi n\ddef g(E)\cr
&=\,\int_{r_1}^L\dd r\g44^{-1}\int(2\ell+1)\dd\ell\sqrt{E^2-\g44\Big(m^2+{
\ell(\ell+1)\over r^2}\Big)}\,,\cr}\eqno(8.11)$$
where the $\ell$-integration goes over those values of $\ell$ for which the
argument of the square root is positive.

To determine the thermodynamic properties of the system, we proceed with
second quantization. Every energy level determined above may be 
occupied by any non-negative number of quanta. The free energy $F$ at an
inverse temperature $\b$ is given by
$$e^{-\b F}\,=\,\sum e^{-\b E}\,=\,\prod_{i=1}^N\prod_{n, \ell,\ell_3}
{1\over 1-\ex{-\b E(n,\ell,\ell_3)}}\,.\eqno(8.12)$$
{}From this one computes
$$\eqalignno{\b F\,&=\,N\sum_\n \log\big(1-e^{-\b E}\big)\,;&(8.13)\cr
\pi\b F\,&=\,N\int\dd g(E)\log\big(1-e^{-\b F}\big)\,=\,-N\int_0^\infty\dd E
{\b g(E)\over e^{\b E}-1}&\cr
&=\,-\b N\int_0^\infty\dd E\int_{r_1}^L\dd r\g44^{-1}\int\big(2\ell+1\big)
\dd\ell&(8.14)\cr
&\times(e^{\b E}-1\big)^{-1}\sqrt{E^2-\g44\Big(m^2+{\ell(\ell+1)\over r^2}\Big)}
\qquad(r_1=2M+h)\,.&\cr}$$
Again the integral is taken only over those values for which the square root 
exists. $N$ is the multiplicity of the fields $\F_i$. In the approximation
$$m^2\,\ll\,2M/\b^2 h\,,\qquad L\,\gg\,2M\,,\eqno(8.15)$$
the main contributions to this integral are found to be
$$F\,\approx\,-{2\pi^3 N\over 45h}\Big({2M\over\b}\Big)^4-{2\over 9\pi}L^3 N
\int_m^\infty{\dd E(E^2-m^2)^{3/2}\over e^{\b E}-1}\,.\eqno(8.16)$$
The second part is the usual contribution from the vacuum surrounding the
black hole at great distances, and as argued before, should be discarded.
The first part is an intrinsic contribution from the horizon, and it is seen
to diverge linearly as $h\downarrow 0$.

The contribution of the horizon to the total energy $U$ and the entropy $S$ 
are $$\eqalignno{U\,=\,{\pa\over\pa\b}(\b F)\,&=\,{2\pi^3\over 15h}\Big(
{2M\over\b}\Big)^4 N\,,&(8.17)\cr
S\,=\,\b(U-F)\,&=\,{8\pi^3\over 45 h}2M\Big({2M\over\b}\Big)^3 N\,.&
(8.18)\cr}$$
When this is adjusted to the Hawking value, Eq.~(7.1), with $\b=1/T\low H=8
\pi M$, we find that the cutoff parameter $h$ must be chosen to be
$$h\,=\,{NG\over 720\pi M}\,.\eqno(8.19)$$
The total energy $U$ of the thermally excited particles is given by
$$G\,U\,=\,\fract38 M\,,\eqno(8.20)$$
independently of $N$. Alternatively, one could have tuned the energy
$U$ to be equal to $m\BH $, which would yield the same order of
magnitude for $h$, but adjusting the physical degrees of freedom, {\it i.e.},
the entropy $S$, appears to us more sensible. Clearly, it makes little sense 
to allow $h{\buildrel{\displaystyle ?}\over\rightarrow}0$, since then both 
the entropy and the energy would diverge.

We refer to the cutoff near the horizon as a ``brick wall". The physical 
distance between the brick wall and the horizon is
$$\int_{r=2M}^{r=2M+h}\dd s\,=\,\int_{2M}^{2M+h}{\dd r\over\sqrt{1-2M/r}}\,=\,
2\sqrt2 Mh\,=\,\sqrt{NG\over 90\pi}\,,\eqno(8.21)$$
which is independent of the mass $m\BH $ of the black hole. The
brick wall should be a property of any horizon of arbitrary size. If $N$ is
not too large, the brick wall thickness is of the order of the Planck length.

The brick wall model, with the values of $\b$ and $h$ fixed according to
Eqs.~(5.40) and (7.1), actually reproduces the thermodynamic properties of
a black hole quite nicely, and could have served as a realistic model for
a black hole that fully obeys Schr\"odinger's equation and preserves
quantum coherence, except for the fact that it also preserves all symmetries 
of the underlying quantum field theory; it could generate chemical potentials 
for the various globally conserved quantum numbers. Thus, not only the
temperature must be constrained to keep the Hawking value, but also the
chemical potentials are constrained to be zero. In principle, this is easy to
realize, simply by introducing symmetry breaking effects in the brick wall
boundary condition, but probably one would then be pushing this model too far;
anyway, its most important deficiency is that we completely gave up invariance
under general coordinate transformations near the horizon.

The most important lesson to be learned from the brick wall model is that 
Hawking radiation can indeed be seen to be compatible with quantum mechanical 
purity, if only one could introduce a cutoff at the Planck scale.

\secbreak
\ni{\bf 9. QUANTUM COHERENCE}\medskip
The field equation (8.6) for non-interacting scalar fields in the
Schwarzschild metric shows an important fact: {\it in terms of the coordinates
$\s$ and $t$ the solutions form plane waves as $\s\rightarrow -\infty$}, or,
the region near the horizon is not compact. If the brick wall were absent, the
entropy of the quanta in this region would be unbounded. But it is equally
evident that, with or without brick wall, it would be incorrect to ignore the
self-interactions. At $\s\rightarrow-\infty$, the plane waves enter an
infinitely inflated region of space-time. The effective Newton constant $G$
becomes infinitely strong here, and therefore any decomposition of Hilbert
space in terms of mutually non-interacting field quanta will be hopelessly
inadequate in this region. Indeed, one possible physical interpretation of the
brick wall model could be that the excessively strong gravitational self
interactions among the field quanta effectively form a barrier at $r<2M+h$,
crudely represented here as a brick wall.

In the derivation of the Hawking spectrum, these gravitational
self-interactions were ignored.  Now this could be partly justified. The
general coordinate transformation which yielded this result has been performed
in the region where the outgoing particles were already fairly soft, so that
indeed ignoring the gravitational self-interactions {\it was} justified, if
our only aim had been to compute the intensity of this radiation. What is {\it
not} justified, is ignoring the gravitational self-interactions when we
compute the {\it reaction of the quantum state of the outgoing matter when new
particles are sent in}.

Consider the Kruskal coordinates described in Section~2. Let an ingoing
particle be at the position $r\approx 2.56~M$ at time $t_0=0$. From Eqs (2.3) we
read off that at this point $xy\approx 1$ and $x/y=1$, and the particle
follows the line $x\approx 1$. Let an outgoing particle arrive at $r\approx
2.56~M$, $xy=1$ at time $t=t_1$. At that time, $x/y =  e^{t_1/2M}$, and this
particle has moved along the line $y= e^{-t_1/4M}$. The two particles have met
each other at time $t=\half t_1$, at the point $xy= e^{-t_1/4M}$, or
$$r\,=\,2M+2M e^{-t_1/4M-1}\,;\qquad t=\half t_1\,.\eqno(9.1)$$
Apparently, if $t_1\gg 4M$, this meeting point is far inside the
trans-Planckian region, and it may not be generally correct to regard the
outgoing particle as a Hawking particle that was not affected by the ingoing
object.

Whether or not this is correct depends on how the outgoing particle is
represented. The Rindler coordinate transformation yields a state for the
outgoing particle that is a thermodynamic mixture of many different quantum
states. It is this mixture which appears hardly to be affected by ingoing
material. Our central assumption, however, is that this thermodynamically
mixed state is merely a {\it macroscopic\/} description, comparable to the way
one describes a vessel containing a gas consiting of a large number of
molecules; but the same vessel also allows for a {\it microscopic\/}
description, in terms of one quantum state only. Similarly, we expect a pure
state description of the outgoing Hawking particles to be possible as well,
but it is these individual states that will be strongly affected by the ingoing
material.

Interactions between in- and outgoing material of course arise due to
gauge theoretic forces as described in the Standard Model of the
elementary particles, but by far the most important force here will be
the gravitational one, whose strength increases without bound in the
trans-Planckian region as $\s\rightarrow -\infty$. The main ingredient
of this gravitational interaction may be conveniently described as a
{\it shift\/} in either the $x$- or the $y$ coordinate. In Fig.~3, the
shift is illustrated. Here, the $x$, $y$ coordinates are chosen such
that the outgoing particle is far on its way out, so the exit time
$t_1$ has been tuned to be close to zero. The ingoing particle has
entered at time $t_0=-\L$, a large negative number.  Consequently, the
ingoing particle has been boosted to such high energies that it
produces a non negligible gravitational field.  The effect this field
has on nearby geodesics is in first approximation a shift, as
indicated, with 
$$\d y(\th,\f)\,\approx\,f(\th,\f,\th',\f')p\in(\th',\f')\,,\eqno(9.2)$$
where $(\th',\f')$ is the angular point where the ingoing particle
crosses the horizon, and $(\th,\f)$ is the intersection of the
geodesic with the horizon. Here, $f$ is a Green function, to be further
described in the next Section. $p\in$ is the momentum in terms of the
locally flat coordinates $(X,Y,Z,T)$ of Eq.~(2.6).

\midinsert\epsffile{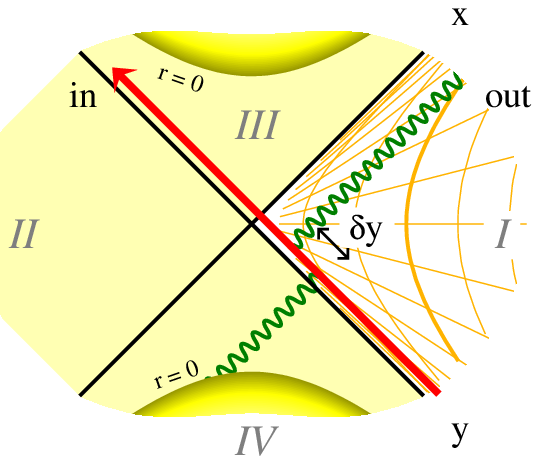}
\narrower\scrunch
\ni Figure 3. The shift in the {\smallit y} coordinate, caused
by an early ingoing particle. If the outgoing particle is soft, the shift it
causes in the {\smallit x} coordinate is negligible.\endinsert

This effect opens the possibility to describe the response of a black
hole to infalling matter in terms of a unitary scattering matrix. We
refer to this theory as the {\it $S$-matrix Ansatz\/}. According to
this theory the outgoing particles do not form a mixture of different
quantum states, but {\it just one state}, which only appears to behave
like a mixed state as soon as we neglect to hand in all existing
information about the ingoing particles. Averaging over the various
possible modes for the ingoing particles will give us back the mixed
state of the outgoing objects, fully in agreement with the calculations
of Section~5.

Naturally, an objection often heard is that if ingoing particles are
considered that are either very energetic, or merely went in
sufficiently long ago, the shift $\d y$ will be so big that a particle
originally on its way out (towards region $I$ in Fig.~3), will be sent
back into the black hole, in region $III$. One could argue that we may
start chosing from many different initial states (where the outgoing
particles were different species, for instance), which would then all
turn into the same state after the shift.

The way to avoid this objection is to assume that the dimensionality of
the Hilbert space of all possible out-states is not that great at all.
We could begin with an ``ideal" black hole, for instance a perfectly
spherical symmetric one, or else, more realistically, with one
particular choice of initial state. This initial black hole produces
Hawking particles {\it in one quantum state only}. All other black hole
states are then obtained from this one by adding or subtracting ingoing
particles. Their effect on the outgoing state is that this state is
deformed by shifts $\d y$. We must insist that there can be no way to
send particles into the hole, or even some arbitrary form of
information, without effecting the outgoing particles by a shift
somewhere along the horizon, or by some other detectable change.

As will be shown in subsequent chapters, one can construct a unitary
scattering matrix along these lines, but a somewhat curious consequence
of this scenario can already be seen here: if the gravitational shift
were taken to be the {\it only\/} possible interaction among particles,
then there could not exist two states that are mutually independent
(orthogonal) and yet identical to each other at all $y\ge\d y$, where
$\d y$ can be arbitrarily large -- if there were, then one could
obliterate this information by sending an extra particle inwards, and
arrive at a contradiction with unitarity. This situation is certainly
quite different from any of the conventional descriptions of quantum
field theory in terms of a Fock space. The existence of
non-gravitational interactions, and more complicated gravitational ones
as well, will give rise to complications in this simple constraint on
the states (of outgoing particles), but since these other interactions
may often be taken to be weak, we expect that indeed in the real world
there are constraints of this remarkable nature.

Another objection often heard is the question of the quantum mechanical
copying machine. Imagine some form of information being sent into the
black hole. An observer in region $III$ (see Fig.~3) could pick up this
information.  But if, as our theory asserts, also an observer in region
$I$, at very late times, can read off this information, since the
ingoing and outgoing states are connected by a unitary $S$-matrix, then
information concerning the quantum phases of wave packets can be
extracted at two places, as if a photo-copying machine had reproduced
it. According to quantum mechanics, this leads to conflicts as soon as
interference experiments are considered. A quantum photo-copying
machine cannot exist. One could add to this that the two observers
under consideration seem to be spacelike separated, hence the operators
they have to their disposal appear to commute.

The answer\ref{25} to this ``objection" is most clearly formulated in
terms of Heisenberg operators (space-time dependent operators acting on
states that do not depend on time). An observer can only move towards
region $III$ if he does not encounter extremely energetic particles
along the $x$-axis while he goes there. This implies that the state of
outgoing particles must be chosen to be exactly the one corresponding
to the given past history of the black hole. Now suppose that a late
Hawking observer switches on his operator in order to measure
something. If this operator has any effect at all on the states he is
considering, he replaces the state of outgoing Hawking particles by
another state. This other state differs from the previous one by having
a given set of (or superposition of) particles now emerging from the
black hole.

The first observer who needed to enter into region $III$ to do his
observation, will be impeded in this act by the energetic particles now
present on the $x$-axis. These particles are so energetic (in terms of
the appropriate coordinates) that region $III$ is essentially blocked
out for him. If he does try to enter anyway, he will find a world quite
different from what he would have seen if the Hawking observer had not
used his operator. All of this merely implies that the operators of
observers in region $III$ and those of Hawking observers do not commute
at all. Their commutators diverge exponentially, as the time difference
increases.

The fact that operators that appear to commute in flat space-time, now
cease to commute in terms of the Rindler variables implies that the
Rindler coordinate transformation must be more complicated than usually
considered. States naturally present in Rindler space (such as the ones
containing arbitrary choices of Hawking particles or arbitrary choices
of early ingoing particles), cannot be accepted in flat space because
of the extreme gravitational curvature they would tend to produce.
Conversely, in flat space, one would expect mutually orthogonal and
independent modes in regions $I$ and $III$ which cannot be chosen
independently as far as the Rindler observer is concerned. We return 
to this issue in Section~16.

The conclusion of this section is that assuming the survival of quantum
mechanical purity in a black hole has far-reaching consequences for the
nature of Hilbert space, even in flat space-time, and that the way the
states transform in a transformation towards Rindler space must then be
far more complex than the linearised procedures usually considered.
This is actually what our theory has in common with string theory.
String theories now also appear to favor the idea that black holes may
form pure quantum mechanical states\ref{14}.

\secbreak
\ni{\bf  10. THE AICHELBURG-SEXL METRIC NEAR A BLACK HOLE}\medskip
The gravitational effect of an infalling particle in the Schwarzschild
metric can be understood when we transform to a locally flat
space-time, Eqs.~(2.6). Consider the coordinate frames of Section~2. As
Schwarzschild time $t$, or equivalently, Rindler time $\t$, evolves,
the infalling particle is Lorentz boosted, as we see in Eq.~(2.8).  In
terms of the flat coordinates, therefore, the energy of the particles
increases exponentially, and thus it quickly reaches values where
gravitational effects can no longer be ignored. These effects are easy
to calculate in the approximation that the source particle moves with
the speed of light\ref{26, 27}.

First consider the case that the surrounding space-time is completely
flat. In the rest frame we can approximate the metric as

$$\dd s^2\,=\,\dd x^2+{2\m \over r}\dd t^2+{2\m \over r}\dd r^2\,,\qquad
r\,=\,\sqrt{x_1^2+x_2^2+x_3^2}\,,\eqno(10.1)$$
where $\m=Gm$, and $m$ is the mass of the source particle. This we rewrite as
$$\dd s^2\,=\,\dd x^2+{2\m \over r}(u\cdot\dd x)^2+{2\m \over r}\dd r^2\,,\qquad
r\,=\,\sqrt{x^2+(u\cdot x)^2}\,,\eqno(10.2)$$
where $$u\,=\,(0,0,0,i)\ ;\qquad u^2\,=\,-1\,.\eqno(10.3)$$
In these expressions, we have neglected all effects that are of higher
order in the particle mass $\m$, since we choose $\m$ to be small. The
particle's Schwarzschild radius $r_+$ is very small, and the Lorentz
boost to be considered next will only further reduce the particle's
dimensions.

The advantage of the notation chosen in Eq.~(10.2) is of course that
now the Lorentz boost is straightforward. In the boosted frame we can
take
$$m\,u^\m\,\Rightarrow\,(0,0,p,ip)\,=\,p^\m\ ,\qquad 
Gp\,=\,{\m v\over\sqrt{1-v^2/c^2}}\,\gg\,\m \,. \eqno(10.4)$$
In the limit $\m\Rightarrow0$, $p$ fixed, one has $r\Rightarrow|x\cdot u|$.
It will turn out to be useful to compare the metric then obtained with
the flat space-time metric in two coordinate frames $y^\m_{(\pm)}$,
defined as
$$y^\m_{(\pm)}\,=\,x^\m\pm  2\m \,u^\m\log r\,.\eqno(10.5)$$
We have:
$$\dd y_{(\pm)}^2\,=\,\dd x^2\pm{4\m \over r}(u\cdot\dd x)\,\dd r
-4\m ^2\,{\dd r^2\over r^2}\,;\eqno(10.6)$$
$$\dd s^2-\dd y_{(\pm)}^2\,=\,{2\m \over r}\,\dd\bigl[r\mp(u\cdot x)\bigr]^2+ 
4\m ^2(\dd\log r)^2\,.\eqno(10.7\pm)$$
 
Now consider the limit (10.4). We keep $p$ fixed but let $\m$ tend to
zero.  We now claim that when $(p\cdot x)>0$, the metric $\dd s^2$
approaches the flat metric $\dd y_{(+)}^2$, whereas when $(p\cdot
x)<0$, we have $\dd s^2\Rightarrow\dd y_{(-)}^2$, and at the plane defined
by $(p\cdot x)=0$ these two flat space-times are glued together according
to
$$y_{(+)}^\m\,=\,y_{(-)}^\m+4\m \,u^\m\log|\tl x|\,,\eqno(10.8)$$
where $\tl x=(X,Y,0,0)$, the transverse part of the coordinates $y^\m$.

Verifying the flatness of space-time away from the plane $(p\cdot x)=0$, is
easy, but to ascertain the connection formula (10.8), is a bit more delicate
(if in Eq.~(10.5), $r$ were replaced by $|(u\cdot x)|$, we would also obtain
flat regions of space-time, but not the connection formula). The complication
is that, though $u^\m\to\infty$, the inner product $(u\cdot x)$ can be small.
Our argument is as follows. First we note that the range of $\log r$ in
$(10.7\pm)$ will diverge not worse than logarithmically in the limit, so when
$\m^2\to0$ we can ignore this last term.  Next, given a number $\l$, we divide
space-time in three regions:
$$\eqalign{A)\qquad &(u\cdot x)\,>\,\l\,;\cr
B)\qquad &(u\cdot x)\,<-\l\,.\cr\hbox{and}\quad
C)\qquad &\left|(u\cdot x)\right|\,\le\,\l\,;\cr }\eqno(10.9)$$
In region $(A)$, we use
$$r-(u\cdot x)\,=\,{x^2\over r+(u\cdot x)}\,,\eqno(10.10)$$
which is therefore bounded by $x^2/\l$. Thus, the first term in Eq.~(10.7+)
is bounded by $\m/\l^2$ times a coordinate dependent function (note that 
$r\ge|\tl x|\,$). Similarly, in region $(B)$, Eq.~(10.7--) will tend to
zero as $\m/\l^2$.
In the region $(C)$, we have that $r$ and $(u\cdot x)$ are both bounded by terms 
that are finite or proportional to $\l$. So, in $(C)$, both $(10.7\pm)$ are
bounded by functions of the form $\m$ or $\m\l^2$. Choosing $\l$ such that, as
$\m\to 0$, both $\m\l^2\to0$ and $\m/\l^2\to 0$, allows us to conclude that
$$\eqalignno{\dd s^2\,&\ra \dd y_{(+)}^2\qquad\hbox{if}\quad (p\cdot x)
\,\gap\, 0\,,&(10.11A)\cr
\dd s^2\,&\ra \dd y_{(-)}^2\qquad\hbox{if}\quad (p\cdot x)\,\lap\, 0\,,&(10.11B)\cr
y_{(+)}\,&=\,y_{(-)}+4\m \,u^\m\log r\qquad\hbox{in the region}\quad (p\cdot 
x)\approx 0\,,&(10.11C)\cr}$$
which is equivalent to Eq.~(10.8). This defines the Aichelburg-Sexl metric\ref{26}.
 
\midinsert\epsffile{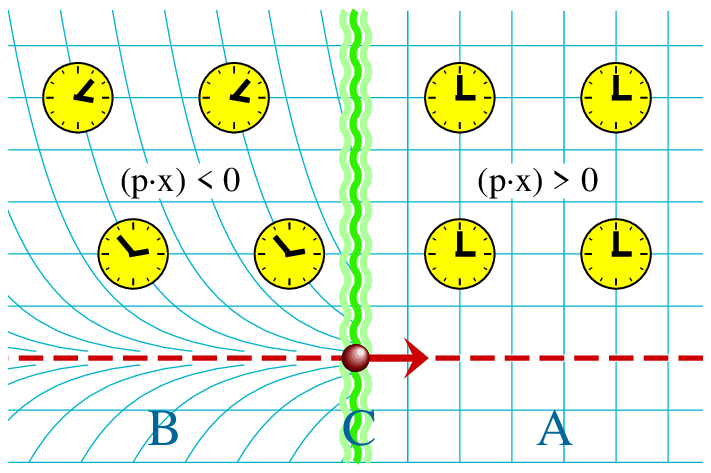}\narrower\scrunch\ni Figure 4. Snapshot of the
gravitational shock wave caused by a highly relativistic particle. If we have
a rectangular grid and synchronized clocks before the particle passed by
(region A), then, behind the particle (region B), the grid will be deformed,
and the clocks desynchronized. The shift is proportional to the logarithm of
the transverse distance. \endinsert
 
The effect a fast moving particle has on the surrounding space-time, is
visualized in Fig.~4. Acually, the picture should be a bit more
complicated since the clocks will also begin to {\it move} after the shock
wave passed. The artist stopped them by giving  them a little kick to
keep them in a fixed position afterwards.

In terms of the light cone coordinates of Eq.~(4.2) in section~4, we have
the connection formula
$$\eqalign{{x^+}_{(+)}-{x^+}_{(-)}\,&=\,4Gp^+\log |\tl x|\,=\,4\sqrt2\, 
Gp\log|\tl x|\,;\cr
{x^-}_{(+)}-{x^-}_{(-)}\,&=\,0\,.\cr}\eqno(10.12)$$
Here, $\tl x$ is the transverse distance from the source particle, which
is moving (highly relativistically) along the line $x^- = \tl x = 0$.
The r.h.s. of Eq.~(10.12) is a Green function,
$-\sqrt2 f(\tl x)p$, satisfying the equation
$$\tds f(\tl x)\,=\,-8\pi G \,\d^2(\tl x)\,,\eqno(10.13)$$
where the sign is chosen such that $f$ is large for small values of
$\tl x$.

Next, we generalize this result for a particle moving into a black hole.
For the Schwarzschild observer, its energy is taken to be so small that
its gravitational field appears to be negligible. However, in Kruskal
space, Section~2, the energy is seen to grow exponentially as Schwarzschild
time $t$ progresses. Let us therefore choose the Kruskal coordinate frame
such that the particle came in at large negative $t$. This means (Eq.~(2.3)),
that $x\approx0$, or, the particle moves in along the past horizon.
In view of the result derived above one can guess in which way the ingoing
particle will deform the metric: we cut Kruskal space in halves across the
$x$-axis, and glue the pieces together again after a shift, defined by

$$y_{(+)}\,=\,y_{(-)}-F(y_{(-)},\th,\f)\,,\eqno(10.14)$$
where $(\pm)$ now refers to the regions $x \glt 0$. 
This corresponds to a metric with a delta-distributed Riemann curvature
on the plane $x=0$. The function $F$ is yet to be determined. Since the 
metric is
$$\dd s^2\,=\,2A(x,y_{(\pm)})\dd x\dd y_{(\pm)}+r^2(x,y_{(\pm)})\dd\W^2
\qquad\hbox{in the regions}\quad x\glt 0\,,\eqno(10.15)$$
with $A(x,y)=(16M^3/r)e^{-r(x,y)/2M}$, and $r(x,y)$ as in Eq.~(2.3), 
one may be tempted to write  
$$\eqalign{y&\ddef y_{(+)}\Th(x)+y_{(-)}\Th(-x)\,;\cr
\dd s^2&\,=\,2A(x,y)\dd x\bigl(\dd y -\d(x)F(y,\th,\f)\dd x\bigr)+
r^2(x,y)\dd \W^2\,,\cr}\eqno(10.16)$$
but some caution is not out of place: if $F$ depends on $y$ this expression
is not unambiguous. We will see that, fortunately, $F$ does not depend on
$y$, but it is always better to refer to Eqs.~(10.14) and (10.15) whenever
the delta function in (10.16) causes difficulties.

The metric described by Eq.~(10.16), and the effect it has on one of the
geodesics, are sketched in Fig.~3. The shift $\d y$ in the $y$-coordinate is
our function $F$.

Let us reserve indices from the beginning of the alphabet ($a$, $b$, \dots)
for indicating the transverse directions $\th$ and $\f$.
The Ricci curvature has been computed in Ref\ref{ 27}. Making use of the fact that,
on the horizon $x=0$, or $r=2M$, the functions $A$ and $r^2$ do not depend
on the coordinate $y$, one obtains
$$\eqalignno{R_{yy} \,&=\,0\,,\qquad R_{ya}\,=\,0\,,&(10.17a)\cr
R_{ab}\,&=\,-h_{ab}{(r^2)_{,yy}\over A}\,F\d\,=\,0\,,&(10.17b)\cr
R_{xy}\,&=\,\left({{A_{,y}}^2\over A^2}-{A_{,yy}\over A}-{2r_{,y}
A_{,y}\over rA}\right)F\d\,=\,0\,,&(10.17c)\cr
R_{xa}\,&=\,-{A_{,y}\over A}F_{,a}\d +\OO[A_{,y},\,A_{,yy}]F_{,a}F\th\d\,=\,0\,,
&(10.17d)\cr
R_{xx}\,&=\,{A\over r^2}\tds F\d-{2r_{xy}\over r}F\d
+\OO\big(A_{,y},\,r_{,y}\big)\big(\d,\,\d_{,x}\big)\,.&(10.17e)\cr}$$
Here, $h_{ab}$ stands for the spherical metric in $(\th,\,\f)$ space, 
$\d$ stands for $\d(x)$, and $\{\ \}_{,y}$ stands for differentiation 
with respect to $y$. Furthermore, $\tds$ stands for the angular Laplacian in the
$\th,\f$ variables.   The last terms in Eqs.~(10.17d) and (10.17e) would 
be ambiguous if $A$ or $r$ depended on $y$. Our construction works
because we take the cut to be along the past horizon, where $x=0$, so
that the $y$ dependence of $A$ and $r$ disappears.

The vacuum Einstein equation would require all $R_{\m\n}$ coefficients
to vanish. Only (10.17e) remains, and putting this equal to zero yields
$$\tds F-{2r\,r_{,xy}\over A}F\,=\,0\qquad\hbox{(vacuum)}\,.\eqno(10.18)$$
Here, the second term can easily be seen to originate from the Riemann
curvature of the $S(2)$ sphere of the $\th,\,\f$ coordinates.

Substituting the known functions $A$ and $r$ we obtain:
$$r_{,xy}\,=\,(4M^2/ r) e^{-r/2M}\,=\, A/4M \,; \qquad
2r\,r_{,xy}\,=\,A\,;\eqno(10.19)$$
$$\tds F-F \,=\,0\qquad\hbox{(vacuum)}\,. \eqno(10.20)$$
In terms of the scaled coordinates $T,X,Y,Z$ of Section~4, the l.h.s. 
of this equation becomes
$$4M^2\tds F-F\,,\eqno(10.21)$$ 
and now we are in a position to insert a source particle.
Let it have momentum $p=p^-/\sqrt2$ in terms of these coordinates (see Eq.~(4.3), 
then combining (10.21) with Eqs.~(10.12) and (10.13), we find in the scaled
coordinates (having the particle enter, for once, at angular coordinates
$\th=\half\pi,\ \f=0$):
$$-\tds F+F/4M^2\,=\,8\pi G{\sqrt{e}\over 4M}(-p^-)\d^2(\tl x)\,=\,
{2\pi G\sqrt{2e}\over M}\,p\,\d^2(\tl x)\,.\eqno(10.22)$$
If we take $p$ to be the momentum with respect to the Kruskal coordinates 
$x$ and $y$, then this equation rescales into:
$$-\tds F+ F\,=\,8\pi G \,(-p_x)\,\d^2(\W)\,,\eqno(10.23)$$
where $p_x$ (which on the mass shell is negative) is the momentum
canonically conjugated to the Kruskal coordinate $x$, and $\W$ is the
angular distance from the ingoing particle.

The solution to the partial differential equation
$$-\tds_\W F(\W,\W')+F(\W,\W')\,=\,\d^2(\W-\W'))\,,\eqno(10.24)$$
for the angular Green function $F$ can be derived by expanding $F$ 
in Legendre polynomials. Taking $\th$ to be the angle between $\W$ and
$\W'$, one finds
$$F(\th)\,=\,(2\pi)^{-1}\sum_\ell {\ell+\half\over \ell(\ell+1)+1}
P_\ell(\cos\th)\,.\eqno(10.25)$$
An integral expression for $F(\th)$ can be found by using the generating
function for the Legendre polynomials,
$$\sum_\ell P_\ell(x)e^{-s\ell}\,=\,(1-2xe^{-s}+e^{-2s})^{-1/2}\,,\eqno(10.26)$$
together with
$$\int_0^\infty e^{-s(\ell+\half)}\cos(\a s)\dd s\,=\,
{\ell+\half\over\ell(\ell+1)+\a^2+\fract14}\,,\eqno(10.27)$$
so that one can rewrite Eq.~(10.25) as
$$F(\th)\,=\,{1\over 2\pi\sqrt2}\int_0^\infty{\cos(\half\sqrt3\,s)\over
\sqrt{\cosh s-\cos\th}}\dd s\,.\eqno(10.28)$$
By rotating the integration contour in the complex plane one can obtain
an integral of a function with a constant sign:
$$F(\th)\,=\,{1\over 2\pi\sqrt2\cosh(\half\pi\sqrt3)}\int_0^{\pi-\th}{\dd\w
\cosh(\half\w\sqrt3)\over\sqrt{\cos\th+\cos\w}}\,.\eqno(10.29)$$
From this expression it is easy to deduce that $F>0$ for all $\th$, and
that
$$F(\pi)\,=\,{1\over 2\cosh(\half\pi\sqrt3)}\,.\eqno(10.30)$$
One may also verify that Eqs~(10.28) and (10.29) have the proper limiting behavior
at small $\th$:
$$F(\th)\,\ra\,(1/2\pi)\log (1/\th)\qquad \hbox{as}\quad \th\,\ra\,0\,.\eqno(10.31)$$

\secbreak
{\bf 11. THE $S$-MATRIX}\medskip
\ni As stated in the Introduction, the postulate that scattering of
particles against a black hole can be described by a quantum mechanical
scattering matrix is an {\it assumption} that cannot be proved from the
principles of quantum field theory and general relativity alone.
Indeed, it may well be at variance with these theories, if the latter
would be extrapolated to beyond the Planck scale. The $S$-matrix Ansatz
applied here may be seen as a new physical principle, perhaps
comparable to Max Planck's new postulate in his 1900 paper, that
energies are quantized. The $S$-matrix Ansatz reads as
follows\ref{28}:\smallskip {\it\narrower All physical interaction processes
that begin and end with free, stable particles moving far apart in an
asymptotically flat space-time, therefore also all those that involve
the creation and subsequent evaporation of a black hole, can be
described by one scattering matrix $S$ relating the asymptotic outgoing
states $|\Out\ket$ to the ingoing states $|\In\ket$.\smallskip}

\ni In essence, the Ansatz will be used in the following way\ref{10,
28}: consider one state $|\In\low0\ket$ and one state $|\Out\low0\ket$,
with a possible black hole in their connecting history. We assume some
value for the transition amplitude $\bra \Out\low0|\In\low0\ket=\NN$.
This means that we replaced the out-state produced by the
Hartle-Hawking vacuum, which actually was a quantum mechanical mixture
of states, by one arbitrary choice $|\Out\low0\ket$. Then, using all
the physical laws that we know and trust, we compute neighboring
$S$-matrix elements, $\bra\Out\low0+\d\out|\In\low0+\d\in\ket$.

If there were no interactions, the effects from $\d\in$ onto the
out-states would not have been discernable. All amplitudes would have
to be equal, and the scattering matrix thus obtained could never be
unitary. Since in the calculations of Section~5, interactions between
the $\F$~particles were ignored, those calculations were not good
enough to give us our $S$-matrix. In this section, we will take only one
type of interaction into account: the gravitational shift computed in
the previous section. Thus, we only consider particles moving in and
out in the longitudinal direction, with hyper-relativistic speeds
when they are near the horizon. Far away from the horizon, as soon
as $r-2M=\OO(2M)$, they will be allowed to go slower, indeed, out-moving
particles may turn around to fall back in again. What has to be done 
in order to accommodate for such possibilities, is to define the 
$S$-matrix to consist of three ingredients:
$$S\,=\,S\out S\low{\rm hor}S\in\,,\eqno(11.1)$$
where $S\in$ relates the asymptotic in-states to wave packets moving inwards
very near the horizon, $S\out$ connects wave packets moving outwards very near
the horizon to the asymptotic out-states, and $S\low{\rm hor}$ is the really
important part telling us how particles moving inwards very near the horizon
affect the outgoing particles very near the horizon.\fn*{This phrazing is more
precise than the ones in previous publications\ref{28}, but it is only partly an
answer to recently uttered criticisms\ref{29}. The splitting of the wave
functions into ingoing and outgoing ones at finite distance from the horizon
will continue to be problematic.} $S\in$ and $S\out$ follow unambiguously from
known laws of low-energy physics, and require little discussion.

The splitting (11.1) allows us to consider the limit $M\to\infty$
without appreciable complications. It is in this limit that the most
important paradoxes arise; indeed, it is the limit that has {\it not}
yet been addressed in terms of string theories.\fnd{In string
theories\ref{14}, only near-extreme black holes can be handled, in which
the horizon has a very tiny area or is degenerate. What concerns us
most in this paper, is the non-degenerate horizon.} In this limit, the
region near the horizon can be described as a Rindler space. The angles
$\th$ and $\f$ are replaced by flat transverse coordinates, and after
rescaling of the momentum $p$, the term $F/4M^2$ in Eq.~(10.22)
disappears. We recover the shift~(10.12), determined by the Green
function $f$ of Eq.~(10.13).

To begin our construction of the $S$-matrix, let us take $\d\in$ to be
one extra particle going in with momentum $\d p^-_\In$, at the
transverse position $\tl x'$. Since we use the conventions of
Section~4, the value of $\d p^-_\In$ is negative.  The outgoing particles,
at points $\tl x$ near to the point $\tl x'$, are shifted inwards, so that
$\d x^-_\Out$ is negative, and
$$x^-_\Out\,\to\,x^-_\Out+\d x^-_\Out(\tl x)\ ,\qquad \d x^-_\Out(\tl x)\,=\,f(\tl
x-\tl x') \d p^-_\In\,,\eqno(11.2)$$ 
where $f$ obeys Eq.~(10.13), or, if from now on $8\pi G=1$, 
$$\tds f(\tl x)\,=\,-\d^2(\tl x)\,.\eqno(11.3)$$

We now temporarily suppress the superscripts $\{^\pm\}$, since
the subscripts {\smallrm in} and {\smallrm out} suffice, and
later we want to reintroduce $\{^\pm\}$ with different sign
conventions.
Any outgoing particle has a wave packet $\j$, oscillating as
$e^{ip_\Out x_\Out}$.  With the shift~$\d x\out$, this wave turns into
$$\ex{ip_\Out x_\Out -ip_\Out \d x_\Out}\,=\,\ex{-i\int\dd^2\tl x\bigl[\d x_\Out(\tl
x)\hat P_\Out (\tl x) \bigr]}\ \j\,,\eqno(11.4)$$
where $\hat P_\Out (\tl x)$ is the operator\fndd{Its relation to the
energy momentum tensor is explained in Section 4.} 
that generates a shift at 
transverse position $\tl x$. It is also the total momentum density
of the outgoing particles at transverse position $\tl x$.

Now combining this with Eq.~(11.2), we see that
$$\j\,\Rightarrow\,\j'\,=\,\ex{-i\int\dd^2\tl x\bigl[\d p_\In f(\tl x-\tl x')
\hat P_\Out (\tl x)\big]}\ \j\,.\eqno(11.5)$$
Repeating this many times, adding (or removing) different ingoing particles 
in the in-state, with momenta adding all up to $P_\In (\tl x')$ at the transverse 
position $\tl x'$, we see that the total effect is:
$$\j'\,=\,\ex{-i\int\dd^2\tl x\dd^2\tl x'\bigl[P_\In (\tl x')f(\tl x-\tl x')
\hat P_\Out (\tl x)\big]}\ \j\,.\eqno(11.6)$$
Notice the complete symmetry between in- and outgoing particles. 
$P_\In (\tl x')$ refers to all momenta of particles going in during a 
certain epoch where we have control over the ingoing particles. 
$\hat P_\Out (\tl x)$ refers to all particles seen going out during a 
similar epoch of observations. Before or after these two epochs we
do not have the opportunity to observe or control. The states there are
kept fixed as much as is possible. Of course, both $P_\In $ and $\hat P_\Out $
are operators; from now on we omit the hat ($\hat{}$).

Noting that, according to the result of Section~7, the total number of 
quantum states should be finite, we have reasons to believe that, by
adding or subtracting a sufficient number of particles, we can generate
{\it all} in-states from $|\In\low0\ket$, and for the out-states it is even more
natural to have $P_\Out (\tl x)$ refer to {\it all} outgoing particles.
It is suggested to describe the in- and out states exclusively by
giving the functions $P_\In(\tl x)$ and $P_\Out(\tl x)$. One then obtains
$$\bra \{P_\Out (\tl x)\}|\{P_\In (\tl x)\}\ket\,=\,\NN\exp\bigl[-i\int\dd^2\tl x\,
\dd^2\tl x'P_\In (\tl x')f(\tl x-\tl x')P_\Out (\tl x)\bigr]\,,\eqno(11.7)$$
where $\NN$ is a common normalization factor. The magnitude of this
factor is fixed by requiring $S$ to be unitary; its phase cannot be
determined, but in most cases it will be a freely adjustable parameter
anyway, since our amplitude tends to violate global conservation laws.

This scattering matrix is indeed unitary, if one imposes the inner product
$$\bra \{P_\In  (\tl x)\}|\{{P_\In}'(\tl x)\}\ket\,=\,\NN'\prod_{\tl x}\d\bigl(
P_\In  (\tl x)-{P_\In}' (\tl x)\bigr)\,,\eqno(11.8)$$
for the in-states, with again some normalization parameter $\NN'$, 
and we impose a similar inner product rule for the out-states.

We should hasten to add, that the $S$-matrix (11.7) cannot be the
ultimate result of our theory, since the states $|\{P_{\rm in\atop\rm out}(\tl x)\}\ket$
with the inner product (11.8) form a {\it continuum} of states, and
this is not the result we want. What this really means is that we still
expect some cut-off mechanism when $|\tl x-\tl x'|$ approaches the
Planck length. Indeed, if $|\tl x-\tl x'|$ approaches the Planck
length, our present result is invalid, since then the {\it transverse}
components of the momenta also produce shifts, and those have not been
taken into account. If, however, we limit ourselves to a ``course
grained" description, specifying only features that are large compared
to the Planck length, and if it could indeed be accepted that
restricting oneself to the gravitational interaction forces only (and
of those only the longitudinal ones), is reasonable, then (11.7) seems
to be a reasonable approximation to the $S$-matrix that we are looking
for. In view of this, let us first further analyze what this $S$-matrix
implies.

Consider the Hilbert space of in-states $|\{P_\In  (\tl x)\}\ket$ with inner
product (11.8), and define an operator $U_\In (\tl x)$ that is canonically
conjugated to $P_\In  (\tl x)$:
$$\eqalignno{[P_\In  (\tl x),\,U_\In (\tl x')]\,&=\,-i\d^2(\tl x-\tl x')\,,&(11.9)\cr
[P_\In  (\tl x),\,P_\In  (\tl x')]\,&=\,[U_\In (\tl x),\,U_\In (\tl x')]\,=\,0 &(11.10)\cr}$$
(We regard all these operators as acting on in-states).
The eigenstates of $U_\In (\tl x)$ are the functional Fourier
transforms of the eigenstates $|\{P_\In  (\tl x)\}\ket$ of the operators
$P_\In  $:
$$|\{U_\In (\tl x)\}\ket\,=\,\NN''\int\DD P_\In  \ex{-i\int\dd\tl x\,
P_\In (\tl x)U_\In (\tl x)}|\{P_\In (\tl x)\}\ket\,,\eqno(11.11)$$
where $\NN''$ is again a normalization factor. 

Writing this as $$\bra\{U_\In (\tl x)\}|\{P_\In  (\tl
x)\}\ket\,=\,\NN'''\ex{i\int\dd \tl x\,P_\In (\tl x)U_\In  (\tl
x)}\,,\eqno(11.12)$$ we find that the states $|\{U_\In (\tl x)\}\ket$
can be expressed in terms of the states $|\{P_\Out (\tl x)\}\ket$, by
using Eq.~(11.7). We find:

$$U_\In (\tl x')\,=\,-\int\dd\tl x\,f(\tl x-\tl x')P_\Out (\tl
x)\,,\eqno(11.13)$$ and similarly:
$$U_\Out (\tl x')\,=\,\int\dd\tl x\,f(\tl x-\tl x')P_\In  (\tl
x)\,,\eqno(11.14)$$ where $U_\Out $ is the operator canonically
conjugated to $P_\Out $, since in addition to Eqs.~(11.9) and (11.10)
we have for the out-states:
$$\eqalignno{[P_\Out (\tl x),\,U_\Out (\tl x')]\,&=\,-i\d^2(\tl x-\tl
x')\,,&(11.15)\cr [P_\Out (\tl x),\,P_\Out (\tl x')]\,&=\,[U_\Out (\tl
x),\,U_\Out (\tl x')]\,=\,0 &(11.16)\cr}$$

By virtue of the fact that Eqs~(11.13) and (11.14) relate operators on 
in-states to operators on out-states, we say that these generate
the $S$-matrix. Rewriting the equations as
$$\tds U_\In(\tl x)\,=\, P_\Out(\tl x)\ ,\qquad \tds U_\Out(\tl x)\,=\,-P_\In(\tl x)
\,,\eqno(11.17)$$
underlines the local nature of these equations with respect to the
transverse coordinates $\tl x$. Also:
$$\bra\{U_\Out (\tl x)\}|\{U_\In (\tl x)\}\ket \,=\,\NN''''\exp\big[-i\int\dd^2\tl x
\,\tl\pa U_\Out (\tl x)\cdot\tl\pa U_\In (\tl x)\big]\,.\eqno(11.18)$$
Because of its local nature, this equation may be suspected to be
more elementary than Eq.~(11.7), which was derived earlier.
Combining (11.18) with (11.12) and the analogous inner product
between the $U_\Out $ and $P_\Out $ eigenstates, we rewrite Eq.~(11.7) as
$$\eqalign{&\bra\{P_\Out (\tl x)\}|\{P_\In  (\tl x)\}\ket\,=\,\NN\int\DD
U_\In (\tl x)\int\DD U_\Out (\tl x)\cr\exp\big[& i\int\dd^2\tl x\big\{-\tl\pa
U_\Out (\tl x)\cdot\tl\pa U_\In (\tl x) +P_\In(\tl x)U_\In(\tl x)-P_\Out(\tl
x)U_\Out(\tl x)\big\}\big]\,,\cr}\eqno(11.19) $$
where $\NN$ is again a different but universal normalization factor 
(Henceforth, we write such factors simply as $\NN$.)

Imagine now that both the in- and the out-state can be completely composed of
a finite number, $N=N_\In+N_\Out$, of particles.  Let us denote the momenta of
the ingoing particles as $p^{-,i}\in$, $i=1,\dots,N_\In$, entering at
transverse coordinates $\tl x^i$, and those of the outgoing particles, at
transverse coordinates $\tl x^j$, as $-p^{+,j}\out$, $j=N_\In+1, \dots,N$. The
reason for the minus sign here, is that now the {total} momentum going into
the horizon can be seen as the sum of all 4-vectors $p^\m$ of the in- and
outgoing particles, as it is usually done in field theory. The operators
$U_\Inout$ are put in a Lorentz vector $x^\m$ without sign
changes:
$$x^+\,=\,U_\In\ ,\qquad x^-\,=\,U_\Out\,.\eqno(11.20)$$
Substituting $$P_\In(\tl x)=\sum_ip\in^{-,i}\d^2(\tl x-\tl
x^i)\ ,\qquad P_\Out(\tl x) =-\kern-.7em
\sum_{j=N_\In+1}^Np\out^{+,j}\d^2(\tl x-\tl x^j)\,,\eqno(11.21)$$
one obtains $$\eqalign{&\bra\Out|\In\ket\,=\,\NN\int\DD
x^+(\tl x)\int\DD x^-(\tl x)\cr\exp\big[&i\int\dd^2\tl x\big\{-\half\tl\pa
x^\m(\tl x)\tl\pa x^\m(\tl x) \big\}+i\sum_{i=1}^Np^{\m,i} x^\m(\tl x^i)\big]
\,.\cr}\eqno(11.22) $$
Here, the transverse components of $x^\m$ are not functionally
integrated over; they are the transverse coordinates.
The factor $\half$ compensates for double counting. The contribution
of the transverse components of $x^\m$ to the integrand must be
subtracted, which corresponds to a renormalization of $\NN$.

It is, however, more realistic to put the external particles in wave
functions that are eigenstates of momenta only. Therefore, we must
convolute this expression by transverse wave functions
$e^{i\tl p^i\cdot\tl x^i}$, where the transverse components of the momenta,
$\tl p^i$, must be kept small compared to the Planck energy (Otherwise, 
it would have been illegal to ignore the transverse gravitational 
shifts.) We then obtain
$$\eqalign{&\bra\Out|\In\ket\,=\,\NN\Big(\prod_i\int\dd^2\tl
x^i\Big)\int\DD x^+(\tl x)\int\DD x^-(\tl x)\cr\exp\big[&i\int\dd^2\tl
x\big\{-\half\tl\pa x^\m(\tl x)\cdot\tl\pa x^\m(\tl x)\big\}
+i\sum_{i=1}^Np^{\m,i} x^\m(\tl x^i)\big] \,,\cr}\eqno(11.23) $$
where now the effects of the wave functions are included in the
contributions of the external momenta $p^{i,\m}$ to the `vertex insertions'.

It is here that the striking resemblence to string amplitudes should be
pointed out. We have the string integrand (for closed strings), as well as the
integration over moduli space, which here is formed by the points $\tl x^i$
where the particles cross the horizon. The fact that the action is linearized
is understandable , since all transverse dimensions have been kept large
compared to the longitudinal ones. What is more surprising is the value of the
string constant: it is equal to $i$, in units where $8\pi G=1$.

The way in which here the black hole horizon is identified with a string
worldsheet is sketched in Fig.~5. At $t\ra -\infty$ we have ingoing closed
``strings". Arriving at the horizon these strings exchange a string, whose
world sheet wraps around the horizon exactly once. The edges of the holes 
left behind are the outgoing closed strings. 

\midinsert\epsffile{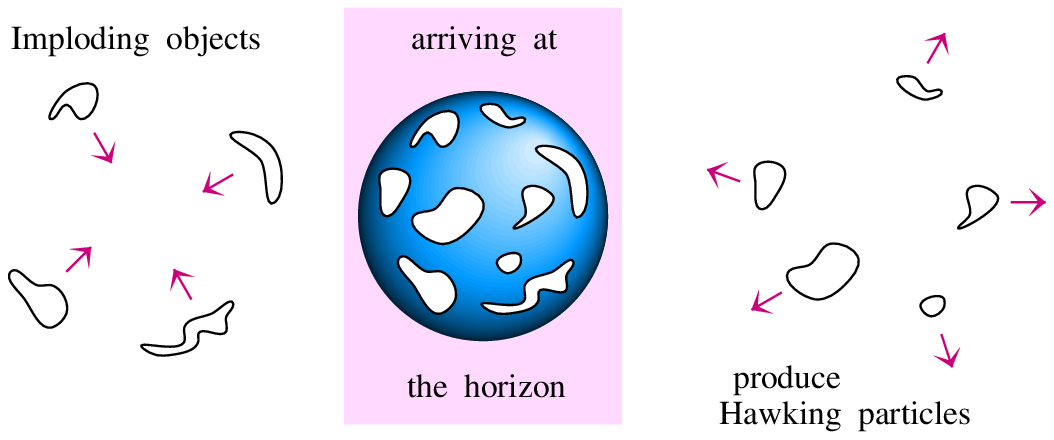}\narrower\scrunch\ni Figure 5. The
horizon as a string world sheet. Three snapshots of a collision event
with a black hole intermediate state.  \endinsert

At this point, let us once again focus on the nature of the Hilbert space
of in- and outgoing particles. Suppose that, for simplicity, we discretize
the transverse coordinates $\tl x$. The functional integrals then become
finite-dimensional. What distinguishes this space from the usual Fock
space is now, that at every point $\tl x$ {\it exactly one} ``particle" is
allowed. The only way to mimick the usual Fock space is to assume
that every elementary point particle must be given a {\it different} value
for its transverse coordinate $\tl x$. This constraint may be considered
to be negligible, if the $\tl x$ are sufficiently fine-grained, but it is
somewhat puzzling how to maintain this constraint in an infinite-volume
limit. Apparently, unlike ordinary Fock space, a state with two or more
particles at the {\it same} transverse position $\tl x$, with momenta
$p^{\m,1}$, \dots, $p^{\m,k}$, is indistinguishable from the state with
just a single particle there, whose momentum is $\sum_{i=1}^kp^{\m,i}$.
This may seem to be odd, but it should be noted that this situation is
identical to what one has in string theory, where the integrand for a
many-particle amplitude is identical to the amplitude for fewer particles,
when two or more of the vertex insertions happen to coincide in the
string world sheet.

The operators $x^\m(\tl x)$ may be regarded as an ``average position"
operator for {\it all} particles ever entering or leaving the horizon at that 
point. This may (partly) explain why this information never ``disappears"
behind the horizon: there are always sufficient numbers of particles
to be seen outside. This is in fact guaranteed by our brick wall model:
the number of particles at distance greater than $h$ from the horizon
are sufficient to represent all ``information" concerning the state of the
black hole.

There is a special interpretation for the commutation rule
$$[U_\Out(\tl x),\,U_\In(\tl x')]\,=\,\int\dd^2\tl x''f(\tl x-\tl x'')
[P_\In(\tl x''),\,U_\In(\tl x')]\,=\,-if(\tl x-\tl x')\,.\eqno(11.24)$$
We could decide to interpret $-U_\Out(\tl x)$ as indicating the position 
of the horizon with respect to the particles seen to emerge from the
black hole, and similarly, $-U_\In(\tl x)$ as the {\it time reverse} of this:
the position of the past horizon with respect to the ingoing particles.
Eq.~(11.24) implies an uncertainty relation for these two quantities.
For ordinary black holes, $U_\Out(\tl x)$ is usually precisely defined,
as it is detemined by the momentum distribution of the ingoing particles
that actually formed the black hole. $U_\In(\tl x)$ is the horizon of the
time-reversed, or ``white hole". In our picture, the white hole is the
object formed by the Hawking particles if we follow these backwards in
time. This is usually spread quantum mechanically over a large range
of values.  In our view, white holes are nothing but quantum super positions 
of black holes. They relate to black holes just like the momentum and
the position of a quantum particle are related to each other.

For future use, we will define the operator $P^\m(\tl x)$ by
$$P^-(\tl x)\,=\,P_\In(\tl x)\ ;\qquad P^+(\tl x)\,=\,-P_\Out(\tl x)\,,
\eqno(11.25)$$
where the minus sign for the outgoing particles is in accordance with 
Feynman diagram conventions in field theory. The transverse components
are still kept small at this stage. Eqs.~(11.17) become
$$\tds x^\m(\tl x) \,=\,-P^\m(\tl x)\,.\eqno(11.26)$$
In addition to Eq.~(11.24) we have 
$$[P^-(\tl x),\,P^+(\tl x')]\,=\,-i\tds\d^2(\tl x-\tl x')\,.\eqno(11.27)$$
 
\secbreak
\ni{\bf 12. ELECTROMAGNETISM}\medskip
It is clear that the $S$-matrix derived in the previous section cannot
be very precise: only the gravitational force was taken into account,
so that it can only be relevant at length scales not too far away from
the Planck scale, whereas beyond the Planck scale the approximation
used again ceases to be valid (the transverse components of the
gravitational force were ignored.) One might try to replace Eq.~(11.23)
by the complete Nambu string action, but the danger then is that this
might be an incorrect way to reproduce Lorentz invariance near the
horizon. Attempts of this nature are postponed to section~15.

In this section we advocate a more cautious and rigorous route towards
improvement, which is the incorporation of other known forces among the
ingoing and outgoing particles. One could include everything known from
the Standard Model, and use more general quantum field theories in
order to conjecture the generally possible expressions for the
$S$-matrix up to details approaching the Planck length. The philosophy
then to be applied is identical to the one used in the previous
sections.

The simplest field theoretical interaction that can be handled is the
electromagnetic one\ref{28}. Consider the effects of electric charge on the
ingoing and outgoing particles. The ingoing particles are not only
characterized by their momentum distribution $P_\In(\tl x)$, but also
by the charge distribution $\r\in(\tl x)$. Computing the
electromagnetic field of a particle entering the black hole along the
past horizon, $x=0$, goes in a way that is similar to the computation
of its gravitational field, Section~10. The field of a static particle
with charge~$Q$  at the origin of a coordinate frame,

$$A_\m(x)\,=\,{Q\over4\pi}{u_\m\over r}\,=\,{(Q/4\pi)u_\m\over\sqrt{
x^2+(u\cdot x)^2}}\,,\eqno(12.1)$$ is Lorentz boosted. One then finds
that, in the limit $u_\pm\ra\infty$, this approaches a pure gauge,
except at the space-time points $x$ satisfying $(u\cdot x)\approx0$,
where the field is singular. To be precise, we again consider the three
regions $(A)$, $(B)$, and $(C)$ of Section~10, Eq.~(10.9).  Let

$$\L(x)\,=\,(Q/2\pi)\log r\,=\,(Q/4\pi)\log[x^2+(u\cdot x)^2]\,,
\eqno(12.2)$$ be a gauge function. We have
$$\pa_\m\L\,=\,{Q\over2\pi}\,{\big(x_\m+u_\m(u\cdot x)\big)\over
x^2+(u\cdot x)^2}\,.\eqno(12.3)$$
In region $(A)$, the vector field $A_\m$ approaches $\half\pa_\m\L$,
and in region $(B)$, it approaches $-\half\pa_\m\L$. As in Section~10,
the most delicate question is, whether in region $(C)$,
$A_\m\pm\half\pa_\m\L$ approaches zero in the sense of distributions.
By carefully comparing Eq.~(12.1) with Eq.~(12.3), one can convince
oneself of this being the case.

We conclude from this calculation, that a charged particle moving with
the speed of light produces an electromagnetic field that takes the
form of a distribution in the transverse plane through the particle.
The regions $(A)$ and $(C)$ are connected to each other by a gauge
rotation $\L =(Q/2\pi)\log|\tl x|$, where $\tl x$ is the transverse
distance from the particle. Again, we can rewrite this as a Green
function,
$$\L(\tl x)\,=\,-f_1(\tl x)Q\ ,\qquad\tds f_1(\tl x)\,
=\,-\d^2(\tl x)\,.\eqno(12.4)$$
On the horizon of the black hole, we must use for electromagnetism
the Green function $f_1(\th,\f)$, obeying
$$\tds_\W f_1(\W,\W')\,=\,-\d^2(\W-\W')\,+\,1/4\pi\,,\eqno(12.5)$$
which differs from the function $F$ used for the gravitational shifts,
by the absence of the second term in Eq.~(10.24). (It is not difficult
to verify Eq.~(12.5) directly from the Maxwell equations in the
Schwarzschild metric; in any case, gauge invariance would forbid any
non-derivative terms in Eq.~(12.5).) The constant $1/4\pi$ is our
temporary resolution to a problem: the integral of the left hand side
over all of the angles $\W$ vanishes, hence the right hand side should
vanish also.  Actually, if the total charge $\int\dd^2\W\,\r\in(\W)$ of
all ingoing particles would not vanish, the result would be a charged
Reissner-Nordstrom black hole, which carries a residual field. we
ignore this field (restricting ourselves to the case when the total
charge is insignificant), and hence we ignore the constant.

In case of a charge distribution $\r\in(\tl x)$ we find the gauge rotation
$$\L(\tl x)\,=\,-\int\dd^2\tl x'\,f_1(\tl x-\tl x')\r\in(\tl
x')\,.\eqno(12.6)$$ Just as in the beginning of Section~11, we
investigate the effect that a small change $\d\r\in(\tl x)$ in the
charge distribution of the ingoing particles has on the outgoing
particles. The wave function $\j(x^-,\tl x)$ of an outgoing particle
with charge $Q_\Out$ is now not only gravitationally shifted, but also
gauge rotated. The gauge rotation is $e^{iQ_\Out\L(\tl x)}$. In total,
if $\r\out(\tl x)$ is the charge distribution of the outgoing
particles, the gauge rotation caused by the ingoing ones is
$$\ex{-i\int\dd^2\tl x\int\dd^2\tl x'\,f_1(\tl x-\tl x')
\r\in(\tl x)\r\out(\tl x')}\,.\eqno(12.7)$$
The sign in this exponent can be verified by comparing with ordinary
Coulomb scattering amplitudes.

We introduce the canonically conjugated operators $\f_\Inout$ as follows:
$$\eqalign{\r_\Inout(\tl x)\dd^2\tl x&\,=\,-i{\pa/\pa\f_\Inout}(\tl
x)\ ;\cr [\r\in(\tl x),\,\f\in(\tl x')]\,=\,-i\d^2(\tl x-\tl
x')\ ;&\qquad [\r\out(\tl x),\, \f\out(\tl x')]\,=\,-i\d^2(\tl x-\tl
x')\ ,\cr\bra\{\f(\tl x)\}|\{\r(\tl x)\}\ket\,=
\,\NN\ex{i\int\dd^2\tl x\,\r(\tl x)\f(\tl x)}&\qquad(\hbox{\it for 
the in- and for the out-operators}).
}\eqno(12.8)$$
Comparing this with Eq.~(12.7), we find
$$\eqalign{\f\in(\tl x)\,&=\,-\int\dd^2\tl x'\,f_1(\tl x-\tl x')
\r\out(\tl x')\,.\cr
\f\out(\tl x)\,&=\,\int\dd^2\tl x'\,f_1(\tl x-\tl x')\r\in(\tl x')\,;\cr
}\eqno(12.9)$$
Consequently, 
$$[\r\in(\tl x),\,\r\out(\tl x')]\,=\,-i\tds\d^2(\tl x-\tl x')\,.
\eqno(12.10)$$

As in Eq.~(12.5), there is a problem with the overall constant; $\f$ is
well-defined up to an overall constant. Changing this overall constant
would be nothing but a global gauge transformation, and is hence
unphysical.  Another important observation is to be made in connection
with the last line of Eq.~(12.8). Since electric charge is quantized in
multiples of the electric charge unit $e$, the charge density
$\r_\Inout(\tl x)$ can only vary by additions of the form $Ne\d^2(\tl
x-\tl x_1)$. This implies that the amplitude~(12.8) cannot change if we
add a multiple of $2\pi/e$ to the field $\f(\tl x)$ (fractions of this
quantum, though also unphysical, do affect the overall amplitude). We
deduce that the model here is actually a sigma model, where $\f$ is
periodic, or rather, the physically relevant field is $e^{2\pi i\f/e}$,
instead of $\f$.

Defining
$$\eqalignno{\r(\tl x)\,&=\,\r\in(\tl x)-\r\out(\tl x)\,;&(12.10)\cr
\f(\tl x)\,&=\,\f\in(\tl x)+\f\out(\tl x)\,,&(12.12)\cr}$$
we have
$$[\f(\tl x),\,\r(\tl x')]\,=\,0\ ;\qquad \f(\tl x)\,=\,\int\dd^2\tl x'
\,f(\tl x-\tl x')\r(\tl x')\ .\eqno(12.13)$$
The gauge amplitude (12.7) can be rewritten as
$$\ex{\half i\int\dd^2\tl x\,\dd^2\tl x'\,f_1(\tl x-\tl x')\r(\tl x)
\r(\tl x')}\,,\eqno(12.14)$$
which differs from Eq.~(12.7) only by two factors that renormalize the
overal phase of the in-state, and the overall phase of the out-state,
but are independent of the way in which these states are put together.

We may rewrite this phase as a functional integral: 
$$\NN\int\DD\f(\tl x)\,\ex{\,i\int\dd^2\tl x\,\big(-\half\tl\pa\f(\tl
x) \cdot\tl\pa\f(\tl x)+\f(\tl x)\r(\tl x)\big)}\,,\eqno(12.15)$$ of
which the classical equation $$\tds\f\,=\,-\r\,,\eqno(12.16)$$
corresponds to Eq.~(12.13).

The expressions of this section must now all be combined with those of
the previous one.  Hilbert space is the product of the two spaces,
being spanned by the states
$$|\{P\in(\tl x)\},\,\{\r\in(\tl x)\}\ket\,\eqno(12.16)$$
All operators of Section~11, referring to momentum and position,
commute with the operators of the present section, referring to electric
charge, and the total amplitude  is given by the direct product of the 
functional integrals (11.19) (or (11.23)) and (12.16).
 
Three remarks are of order:  \item{1.}Apparently, adding the
electromagnetic force makes Hilbert space much larger than it already
was. Actually, a cut-off is needed in any case. The electromagnetic
force gives us no clue as to how this cut-off should be introduced.

\item{2.}In every respect, the electromagnetic charge density resembles
a fifth component of the momentum. The conjugate parameter, $\f$, is a
periodic coordinate. Thus, in our approach, electromagnetism {\it
automatically\/} emerges in the form of a Kaluza-Klein theory.
 
\item{3.}It seems obvious that we can generalize this approach to handle
other gauge forces.  

\secbreak
\ni{\bf 13. OTHER FORCES}\medskip
In the Standard Model of elementary particles, we have non-Abelian
gauge forces, fermions and at least one scalar field.  A complete
implementation of all these variables in the black hole $S$-matrix
appears to be difficult at present, but we will briefly discuss all of
them.

One could introduce the non-Abelian forces directly by adding further
compactified dimensions to space-time. This is probably the procedure
that works best, but it is not quite in agreement with the strategy
proposed all along in this paper. One should first prove that the
physically observed charges, such as weak isospin and $SU(3)$ color,
actually necessitate the introduction of such degrees of freedom. This,
however, is somewhat problematic. If we add a single gauge-colored
charge to the ingoing state, the latter will become gauge
non-invariant, and therefore unphysical. Remember that we had a problem
of this sort even in the Abelian case, but there it could be
controlled. Secondly, non-Abelian charges are not additive, in contrast
to the Abelian ones, so that a semiclassical calculation of their
effects on outgoing particles, as we did this in the Abelian case, is
not possible.  Probably the only thing one can do in order to estimate
the non-Abelian effects on the $S$-matrix, is first to compute the
amplitudes with one-photon exchange between ingoing and outgoing
particles.  The generalization of the single periodic field $\f$ for
the electromagnetic case is then a sigma field $\W(\tl x)$, with
elements in the space of gauge group transformations.  Thus, the
non-Abelian gauge theory generates a sigma model on the two-sphere.

From now on, we refer to the ordinary field theoretical interactions
among elementary particles as ``the Standard Model". What we mean by
this is that we are free to assume any kind of theory that may possibly
be relevant at some energy scale between a GeV and the Planck scale,
and it is assumed to contain gauge vector fields, scalar fields and
spinors.

Suppose we want to repeat the procedure of Section~12 for particles
interacting with a scalar Yukawa field $\f(x)$. The mass $\m$ of this
scalar field may or may not be zero. Using the same notation as in
Eq.~(12.1), an ingoing particle will generate a scalar field of the
form
$$\f(x)\,=\,{g\over 4\pi r}\ex{-\m r}\,=\,{g\over 4\pi}{e^{-\m \sqrt{x^2+
(u\cdot x)^2}}\over \sqrt{x^2+(u\cdot x)^2}}\,,\eqno(13.1)$$
As Schwarzschild time proceeds, the particle's velocity vector $u^\m$
will be boosted more and more towards the past horizon. Unlike
Eq.~(12.1), however, there is no factor $u^\m$ in the numerator, and
while the support of this field becomes more and more flattened against
the past horizon, its intensity does not increase. Outgoing particles
going through this pancake shaped field will be less and less
influenced by it.  As we are mainly interested in the interactions
between ingoing and outgoing particles very near the horizon, where
their relative center of mass energy approaches the Planck energy, the
effects of scalar field exchanges will be extremely small. In
Section~12, we were able to produce a meaningful amplitude in the limit
$u^\m\ra \infty$; in the case of scalar field exchange, this limit does
not produce any non-trivial contribution to the amplitudes.

Yet there is a way in which scalar fields in the Standard Model may
play a role (of course, in a complete theory, without simplifying
approximations, they must be relevant). Assume that 
one or more of the gauge forces in our ``Standard Model" are
condensed into a Higgs mode. This effect may be entirely due to scalar
fields, and yet its consequences for the black hole
$S$-matrix are non-trivial and can be computed. Take the $U(1)$ case. 
The procedure is
exactly as described in the Section~12, except that the Coulomb
field is assumed to have a finite range. For collisions at a very high
center-of-mass energy and a relatively low impact parameter, the only
force felt is the Maxwell field with a Yukawa suppression factor.
Instead of Eq.~(12.1) we have $$A_\m(x)\,=\,{Q\over4\pi}\,{u_\m\over
r}\ex{-m\llow A r}\,,\qquad r=\sqrt{x^2+(u\cdot x)^2}\,,\eqno(13.2)$$ 
where $m_A$ is the photon mass acquired through the Higgs mechanism.
The gauge rotation function $\L(x)$ of Eq.~(12.2) now takes the form
$$\L(x)\,=\,-{Q\over2\pi}\int_r^\infty\dd\s{\ex{-m\llow A\s}\over
\s}\,.\eqno(13.3)$$
With
$$\pa_\m\L\,=\,{Q\over2\pi}\,{x_\m+u_\m(u\cdot x)\over r^2}\ex
{-m\llow A r}\,,\eqno(13.4)$$
we see, as in Section~12, that in regions $(A)$ and $(B)$ we have
$A_\m\Rightarrow +\half\pa_\m\L$ and $-\half\pa_\m\L$, respectively. In
region $(C)$, both $A_\m$ and $\pa_\m\L$ approach to zero in the
distributional sense.  

Thus, as in Section~12, outgoing particles are phase shifted, but now
te phase shift function $\L(\tl x)$ is given by Eq.~(13.3), with
$r=|\tl x|$.  Can we set up an algebra as in Section~12 \thinspace? Not
quite. The function $\L(\tl x)$ does not obey a field equation as
useful as Eq.~(12.5).  The equation it does satisfy is:\fn*{This treatment is
more precise than the original one in ref\ref{ 28}}
$$\Big(\tds+m\llow A{\pa\over\pa r}\Big)\L\,=\,{Q\over2\pi}\,\d^2(\tl x)
\,,\eqno(13.5)$$
where the second term involves the gradient in the radial direction.
This equation does not allow a linear superposition in the case of many
different sources at different locations in the transverse plane.
We do have, for large $r$,
$$\eqalignno{\L(r)\,&\Rightarrow\,-\,{\ex{-m\llow Ar}\over m\llow Ar}
\left(1-{1\over m\llow Ar}+{2!\over(m\llow Ar)^2}-\,\cdots\,\right)\,;
&(13.6a)\cr
{\pa\over\pa r}\L\,&=\,{\ex{-m\llow Ar}\over r}\,,&(13.6b)\cr}$$
and from this it follows that, at sufficiently large $m\llow Ar$, our
field $\L$ tends to obey the Klein-Gordon equation:
$$\big(\tds-m^2\llow {\!A}\,\big)\L(\tl
x)\,\Rightarrow\,{Q\over2\pi}\d^2(\tl x)\,.  \eqno(13.7)$$

{}From this equation it is clear what the main {\it infrared\/} effect
is of a Higgs mechanism in the electromagnetic contribution to our
amplitudes: the Green function $f(\tl x-\tl x')$ obtains a mass term in
its defining equation.  In Eq.~(12.14), the integrand in the exponent
receives a mass term of the form

$$-\half m^2\llow {\!\!A}\,\f^2(\tl x)\,.\eqno(13.8)$$
Here, $m\llow A$ is the gauge photon mass. What is remarkable about
this modification term is that it appears to break the gauge symmetry
$\f\ra\f+\l$. We find that, if in the Standard Model a {\it local\/}
symmetry is {\it spontaneously\/} broken by  Higgs mechanism, then in the
two-dimensional system dscribed in Eq.~(12.14), the corresponding {\it
global\/} symmetry is broken {\it explicitly}.

Clearly, in this particular case, a scalar field can have an effect on
the black hole back reaction amplitude, even if the outgoing particles
emerge at arbitrarily late Schwarzschild time. However, the only visible
effect is a modification of a field equation. This, we believe, is the
only way a fundamental scalar field in the Standard model can betray
its presence in the black hole scattering amplitude: it modifies other
field equations.  The fact that a spontaneous symmetry breaking effect
now turns into an explicit symmetry breaking efect, can be interpreted
as follows. The scalar field under consideration, may have some
accidental value at the point where past horizon and future horizon
meet. In terms of the 4-dimensional theory, a non-vanishing value of
the scalar field there, may imply the spontaneous breakdown of a
symmetry. But for the observer of a black hole, the scalar field is
permanently present at that point.  Its value will not change as the
black hole ages. So, it appears to break the symmetry explicitly.

The next subject to be studied is the question how the horizon
amplitudes are affected if the 4-dimensional theory includes a
confinement mechanism, such as the property that keeps the quarks
together in a hadron. Confinement is often understood as a condensation
of magnetic charges in the theory.  A non-Abelian gauge theory is first
decomposed into a system of photons representing the Cartan subalgebra
of the gauge group, treating the remaining photons together with the
quarks as ordinary charged particles, and the inevitable gauge
singularities one then gets: the magnetic monopoles.  If one then
assumes that these magnetic monopoles condense, one has a satisfactory
description of a quark confinement mechanism\ref{37}. Before discussing this
any further, we must understand how the black hole amplitudes for
states with {\it magnetic\/} charges can be understood.

Consider an operator $\s(\tl x)$ that creates a magnetic monopole
in the in-state at the transverse position $\tl x$. It is not difficult
to describe the electromagnetic field produced by a monopole with a
speed close to that of light. An electric charge produces a Maxwell field,
$$\eqalign{F_{\m\n}=\pa_\m A_\n-\pa_\n A_\m\ ,\qquad\hbox{with}&\quad  
A_+\,=\,\pa_+\big(\half\L(\tl x)
\,{\rm sgn}(x^+)\big)\,,\cr\hbox{or,}\qquad 
F_{+a}^{\,\rm el}\,&=\,\d(x^+)\pa_a\L(\tl x)\,,\cr}\eqno(13.9)$$
where the index $a$ is in the transverse direction, and $\L(\tl x)$ is
given by Eq.~(12.2). The other components of $F_{\m\n}$ vanish. The field 
of a magnetic charge is the dual of this:
$$F_{+a}^{\,\rm magn}\,=\,g\low{+-}\e^{-+ab}F_{+b}^{\,\rm el}\,=\,
-\e_{ab}\d(x^+)\pa_b\L(\tl x)\,=\,\d(x^+)\pa_a\L^D(\tl
x)\,,\eqno(13.10)$$ where $\L^D$ ($D$ stands for ``dual") is a gauge
function that is multiply connected if we rotate around the origin:
$L\ra\L\pm 2\pi/e$. Thus, the operator $\s(\tl x)$ produces a
``frustration" in the field $\f(\tl x)$, a locally stable vortex.  Such
operators are known as ``disorder operators".\ref{36, 37}

We are now in a position to treat the confinement mechanism. The
disorder operators $\s(\tl x)$ that we introduced fail to commute
with the $\f(\tl x')$ fields, as follows:\ref{37}
$$\ex{(2\pi i/e)\f(\tl x)}\,\s(\tl x')\,=\,\s(\tl x')\,\ex{(2\pi
i/e)\f(\tl x)}\, \ex{i\th(\tl x-\tl x')}\,,\eqno(13.11)$$
where $\th(\tl y)$ is the angle of the 2-vector $\tl y$.
\def\table#1{\topinsert$$\vbox{\offinterlineskip\halign{#1}}$$\endinsert}
\def\linespace{height3pt\boxnum\cr} 
\def\topline{\noalign{\smallskip}\noalign{\hrule}\linespace}
\def\midline{\linespace\noalign{\hrule}\linespace}
\def\bottomline{\linespace\noalign{\hrule}\noalign{\smallskip}}
\def\vblock{&\omit&}
\def\boxnum{\vblock\vblock}		 

\table{&\vrule#&\strut\ \ #\hfil\quad\cr  
\multispan{5}\hfil\bf Table 1\hfil\cr 	 
\topline 				 
&\hfil STANDARD MODEL IN && \hfil INDUCED 2 DIM. OPERATOR&\cr
&\hfil 3+1 DIMENSIONS	&& \hfil FIELD ON HORIZON	&\cr
\midline 				 
&\bal \ {\it Spin 2\/}:\hfil $g_{\m\n}({\bf x},t)$ &&\qquad String variables
\ ({\it spin 1\/}): &\cr
&\quad local gauge generator: $u^\m({\bf x},t)$ &&\hfil $x^\m(\W)$\hfil
&\cr \midline
&\bal \  {\it Spin 1\/}:\hfil$A_\m({\bf x},t)$ &&\qquad Scalar variable
\ ({\it spin 0\/}):  &\cr
&\quad local gauge generator: $\L({\bf x},t)$ mod ${2\pi/e}$ && \hfil
$\f(\W)$ mod $2\pi/
 e$\hfil&\cr \midline
&\bal \ {\it Spin 0\/}:\hfil $\f({\bf x},t)$ &&\qquad Coupling constant &\cr
\midline
&\bal \ Magnetic monopole&&\qquad Vortex&\cr\midline
&\bal \ Higgs mechanism: &&\quad Explicit symmetry breaking: &\cr
&\quad``spontaneous" mass $m\llow A$ for vector field &&\quad $\f(\W)$
obtains same mass $m\llow A$. &\cr \midline
&\bal \ Confinement in vector field $A_\m$ &&$\matrix{\hbox{ $\f$ must
be replaced by disorder}\hfill\cr \hbox{operator $\s$; its symmetry
broken.}\cr}$&\cr\midline
&\bal \ Non-Abelian gauge theory &&$\matrix{\hbox{Scalar fields
$W(\W)$, corresponding}\cr \hbox{to non-linear sigma
model.}\hfill\cr}$ &\cr\midline
&\bal \ {\it Spin $\half$}: \ fermions && \qquad Unknown  &\cr \midline
&\bal \ {\it Spin $\fract32$}: \ gravitino &&\qquad Fermion,&\cr
&\quad local gauge generator \ spin $\half$ &&\qquad \it Spin $\half$
&\cr \bottomline}

The disordered phase of a field theory would be characterized by a
spontaneous breakdown of the symmetry
$$\s\ra\s\,e^{i\l}\,,\eqno(13.12)$$ but if confinement is to be
described as the dual analogue of the Higgs mechanism, we must conclude
that the disorder field obtains a mass term, thus becoming a short
range field, and the disorder symmetry (13.12) must be broken {\it
explicitly}.  After such an explicit symmetry breaking it has become
impossible to transform back to the original fields $\f(\tl x)$.

Not much has been done yet to understand the consequences of the
presence of {\it fermions} in the Standard Model. The transformation
properties of a fermion field under Lorentz boosts is intermediate
between those of vectors and those of scalars. Therefore, we expect the
effect of a spinorial field produced by ingoing particles also to scale
away as Schwarzschild time proceeds. 

In {\it supergravity theories}, fundamental fields occur with spin
equal to $\fract32$. Such fields will be enhanced, and therefore play a
more prominent role on the horizon. As is the case for gravity and for
electromagnetism, the {\it generator\/} of a local supersymmetry
transformation will correspond to an operator on the horizon. Thus,
supergravity theories are expected to produce a spin~$\half$ operator
field on the horizon. The conclusions of this Section are summarized in
Table~1.
\secbreak

\ni{\bf 14. THE TRANSVERSE GRAVITATIONAL FORCE}\ref{28}\medskip
As was shown in the previous section, adding the effects from known
(and unknown) quantum field theoretic forces in our calculations for
the scattering matrix, rather adds more states to our Hilbert space
than reducing it. Clearly, we would like to know which of the physical
laws should be taken into account if we want to understand the area --
entropy relation. A promising candidate is the transverse component of
the gravitational shift effect; so-far we have only taken its
longitudinal part into account.

Unfortunately, as we will show, including the transverse gravitational
force is difficult. We here only give an indication as to how one could
proceed along these lines, so as to further improve our theory.

Let us ecapitulate our algebra. From Section~11:
$$\eqalignno{[P_\In(\tl x),\,P_\In(\tl y)] \ =\ 0\ =\ 
&[P_\Out(\tl x),\,P_\Out(\tl y)]\ ;&(14.1)\cr
[U_\In(\tl x),\,U_\In(\tl y)]\ =\ 0\ =\ 
&[U_\Out(\tl x),\,U_\Out(\tl y)]\ ;&(14.2)\cr
[P_\In(\tl x),\,U_\In(\tl y)]\ =\ -i\d^2(\tl x-\tl y&)\ =\ 
[P_\Out(\tl x),\,U_\Out(\tl y)]\ ;&(14.3)\cr
P_\Out(\tl x)\,=\,\tds U_\In(\tl x)\ ;\qquad&P_\In(\tl x)\,=\,-\tds
U_\Out(\tl x) \ ;&(14.4)\cr
[U_\In(\tl x),\,U_\Out(\tl y)]\,=\,if(\tl x-\tl y)\ ;\qquad&
[P_\In(\tl x),\,P_\Out(\tl y)]\,=\,-i\tds\d^2(\tl x-\tl y)\ ,
&(14.5)\cr}$$ and from Section~12:
$$\eqalignno{[\r\in(\tl x),\,\r\in(\tl y)] \ =\ 0\ =\ 
&[\r\out(\tl x),\,\r\out(\tl y)]\ ;&(14.6)\cr
[\f\in(\tl x),\,\f\in(\tl y)]\ =\ 0\ =\ 
&[\f\out(\tl x),\,\f\out(\tl y)]\ ;&(14.7)\cr
[\r\in(\tl x),\,\f\in(\tl y)]\ =\ -i\d^2(\tl x-\tl y&)\ =\ 
[\r\out(\tl x),\,\f\out(\tl y)]\ ;&(14.8)\cr
\r\out(\tl x)\,=\,\tds \f\in(\tl x)\ ;\qquad&\r\in(\tl x)\,=\,-\tds
\f\out(\tl x) \ ;&(14.9)\cr
[\f\in(\tl x),\,\f\out(\tl y)]\,=\,if(\tl x-\tl y)\ ;\qquad&
[\r\in(\tl x),\,\r\out(\tl y)]\,=\,-i\tds\d^2(\tl x-\tl y) 
&(14.10)\cr}$$
(since we neglect the angular curvature of the horizon, we may take 
$f_1$ to be equal to $f$.)

These equations clearly illustrate that charge density may be seen as a
fifth momentum component, and $\f$ as a fifth coordinate.  Now, it is
not difficult to see that what is missing here is the transverse
components, $\tl P(\tl x)$, of the momentum densities. These should obey
similar commutation rules. Indeed, an ingoing particle with large
transverse momentum does produce a shift among the outgoing particles
in the transverse direction.

However, transverse momentum is not an independent variable; it is
related to the $\tl x$-dependence of all other fields. Let $f(\tl x)$
be any operator valued function of the transverse coordinates. The
total transverse momentum operator, $\tl P\tot$, obeys the commutation
rule,
$$[\tl P\tot,\,f(\tl x)]\,=\,-i\tl\pa f(\tl x)\,.\eqno(14.11)$$
It is the integral of a transverse momentum density, $\tl P(\tl x)$.
We distinguish the transverse momentum density of the in-states from
that of the out-states, so
$$[\tl P_\In(\tl x),\,f_\In(\tl y)]\,=\,-i\d^2(\tl x-\tl y)\,\tl\pa 
f_\In(\tl x)\,,\eqno(14.12)$$
and similarly for the out-states. 

For an operator $\r(\tl x)$ that is a
{\it density}, which means that under displacements it transforms with
a Jacobian, such that its integral is conserved, we must demand
$$[\tl P_\In(\tl x),\,\r\in(\tl y)]\,=\,i\r\in(\tl x)\tl\pa\d^2(\tl
x-\tl y)\,.\eqno(14.13)$$

Constructing an operator $\tl P_\In(\tl x)$ with the properties (14.12)
and (14.13) is straightforward; the answer is an elementary exercise in
quantum field theory:
$$\tl P_\In(\tl x)\,=\,P_\In(\tl x)\,\tl\pa U_\In(\tl x)+
\r\in(\tl x)\,\tl\pa \f\in(\tl x)+\cdots\,,\eqno(14.14)$$
where the dots stand for contributions of any other kinds of local
operators fields that may exist in our Hilbert space.
With this definition, the commutation rules~(14.12) and~(14.13) 
follow directly from Eqs.~(14.1--10). We also have 
$$[\tl P^a_\In(\tl x),\,\tl P^b_\In(\tl y)]\,=\,i\tl P^a_\In(\tl y)
\pa_\b\d^2(\tl x-\tl y)+i\tl P^b_\In(\tl x)\pa_a\d^2(\tl x-\tl y)\,.
\eqno(14.15)$$  
The relation between these operators and the energy-momentum density is
explained in Section~4. Latin indices $a$, $b$, \dots, are in the
transverse direction. The tilde ($\tl{\ }$) on $P^a$ is there only to
distinguish this operator from the longitudinal ones. Deriving
Eq.~(14.15) form (14.14) is facilitated by observing the following
important properties of the Dirac delta distribution:
$$\eqalignno{\d^2(\tl x-\tl y)\{f(\tl x)-f(\tl y)\}\,&=\,0\,;
&(14.16)\cr
\pa_a\d^2(\tl x-\tl y)\ \{f(\tl x)-f(\tl y)\}\,+\,\d^2(\tl x-\tl y)&
\{\pa_a f(\tl x)\}\,=\,0\,.&(14.17)\cr}$$
Eq.~(14.17) is, of course, obtained by differentiating (14.16) 
with respect to $\tl x$.
The out-states can be subjected to operators $\tl P^a_\Out$ that obey
an algebra similar to (14.15).

Consider now, as before, our deductive procedure for the production of
scattering matrix amplitudes. This time, we bring about a small change
$\d \tl P_\In(\tl x)$ in the transverse momentum distribution of the
in-state; one of the ingoing particles, for instance, was replaced by 
a particle with a small change in its transverse momentum components.
The gravitational shift, caused by this particle, is now in a slightly
different direction. From Eq. (11.2), we derive that the shift in the 
transverse direction is
$$\d\tl x_\Out\,=\, \int\dd^2\tl x'\,f(\tl x-\tl x')\,\d\tl 
P_\In(\tl x')\,,\eqno(14.18)$$ 
which rotates the phase of the outgoing state as follows:
$$|\j\ket\out\ \ra\ \ex{-i\int\dd^2\tl x\,\dd^2\tl x'\,f(\tl x-
\tl x')\,\d\tl P_\In(\tl x')\cdot\tl P_\Out(\tl x)}\,|\j\ket\out\,.
\eqno(14.19)$$
This is entirely analogous to Eq.~(12.7), and therefore one expects
commutation rules of the form
$$[\tl P^a_\In(\tl x),\,\tl P^b_\Out(\tl x')]\,\qu\,-i\d^{ab}\tds\d^2
(\tl x-\tl x')\,.\eqno(14.20)$$

Unfortunately, Eq.~(14.20) cannot be correct. For one thing, it
violates the Jacobi identities when combined with Eq.~(14.15). In any
case, when Eq.~(14.14) is substituted, much more complicated
expressions for the commutator are obtained. There are several reasons
for this apparent contradiction to arise: Eqs.~(14.4), (14,5), (14.9)
and~(14.10) should be violated due to the transverse shift, and
furthermore, the non-commutativity of $\tl P(\tl x)$ and $P(\tl x')$
should be taken into account.

Presently, it is beyond our capabilities to set up a completely
consistent algebra for the in- and out-states on the horizon. Perhaps,
when a better understanding of string theory is achieved, we might
expect that string theory could provide for a consistent prescription
here.  Conversely, since the first principles from which we arrived at
this point are conceptually very straightforward, one might expect that
these could form a better guideline for further improvement, whereas an
interpretation in terms of strings, as in Section~11, may come
afterwards.

\secbreak

{\ni\bf 15. ALGEBRAS WITH A TRANSVERSE CUT-OFF}\medskip
Qualitatively, it can be understood why the transverse gravitational
force will produce a cut-off at the Planck scale on the horizon.
Imagine a particle going in at transverse position $\tl x$, with
transverse momentum $\tl p\equiv\d\tl P_\In(\tl x)$ whose absolute
value is much larger than the Planck energy. The transverse shift
brought about by this particle is non-negligible, and the out-state is
rotated by the phase factor~(14.19).  Since the shift is large, already
the small values of $\tl P_\Out(\tl x)$ contribute. Thus, in terms of
the out-states, the substitution $\tl P_\In\ra\tl P_\In+\d\tl P_\In$
does not yield particles with very large values of $\tl P_\Out$, and
can therefore be seen as an operation that keeps the state in the
subspace of Hilbert space spanned by the low momentum particles. It is
conceivable, therefore, that states with high transverse momentum can
be formed by linear superposition of states with low transverse
momenta.

It is clear that, when a Planckian resolution is required in the
transverse coordinates, the models of sections~11~--~14 are inadequate.
As long as our understanding of the situation at the Planck length is
limited, we take our resort to some models. Surely, these models will
show deficiencies as well, but they illustrate how discrete physics may
arise naturally, and they may show how to understand the black hole
entropy in terms of the density of states in Hilbert space.

Our first model is obtained by introducing discreteness in the
transverse coordinates by hand, merely by substituting the continuum of
the points $\tl x$ by a lattice. The lattice is chosen to be a random
one, since this allows us to accommodate for curvature on the horizon.
In short: we have points $\{A,\,B,\,\dots\}$, connected by links, see
Fig.~6.

\midinsert\epsffile{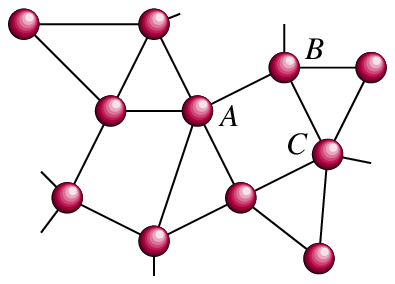}\narrower\scrunch\cl{Figure 6. Latice
points on the horizon, with their link connections.} \endinsert

The ``lengths" of the links are {\it not\/} specified; even if one would
start by postulating them to be of infinitesimal length only, one will
later find the generic separation between neighboring points to be of
the order of the Planck length, as will be demonstrated.  The topology
of the link connections is kept invariant during the
evolution.\fn*{This is a weak point in the model; since we expect no
truly conserved quantities on the black hole, we should expect a
dynamical evolution of the link topology. One may consider further
refinements of this model that take such an evolution into account.} On
every point $A$, we have three coordinate-like degrees of freedom
characterizing the in-state, being the 3-vector:
$${\bf x}_A^\In\,=\,(x_A^{1,\In},x_A^{2,\In},x_A^{3,\In})\,
\ddef\,\big(\tl x\in,\, U\in(\tl x\in)\big) \,,\eqno(15.1)$$
and its conjugated momentum 3-vector:
$${\bf p}_A^\In\,\ddef\,\big(\tl P\in(\tl x\in),\, P\in(\tl
x\in)\big) \,,\eqno(15.2)$$
and similar definitions for the out-states.

The absence of the factors $\sqrt2$ from the lightcone coordinate
definitions (4.2), (4.3) of Section~4, in the third components of
Eqs.~(15.1) and~(15.2), where one would have expected them, is a peculiarity
of this model. The commutation rules are:
$$\eqalign{[x_A^{i,\In},\,x_B^{j,\In}]\,&=\,0\,,\cr
[p_A^{i,\In},\,p_B^{j,\In}]\,&=\,0\,,\cr
[x_A^{i,\In},\,p_B^{j,\In}]\,&=\,i\d^{ij}\d_{AB}\,,\cr}\eqno(15.3)$$
for all $A,\,B$, and where $i,\,j$ take the values 1, 2 and~3.  The
commutation rules for the out-operators are identical.

In order to generate relations between the in- and the
out-operators resembling Eqs.~(14.4), (14.5) and~(14.20), we need a
discrete version of the transverse Laplacian on our lattice. Define
$$\eqalign{C_{AB}\ &=\ \Big\{\matrix{&0\qquad\hbox{if $A$ and $B$ are not 
directly connected,}\cr &1\qquad\hbox{if $A$ and $B$ are linked
neighbors.}\hfill\cr} \cr C_{AA}\ &=\ -N(A)\,,\cr}\eqno(15.4)$$
where $N(A)$ is the number of neighbors linked to the point $A$.
An operator that approaches the Laplacian in the continuum limit
is then given by:
$$\l \sum_BC_{AB}\,F_B\,\Rightarrow\,(\tds F)_A\,,\eqno(15.5)$$
where the constant $\l$ depends on the average number of neighbors
that the points $A$ have, for instance:
$$\l\ =\ \Big\{\matrix{1\qquad\hfill\hbox{for a square lattice;}\hfill\cr
1/\root 4 \of 3\qquad\hbox{for a triangular lattice.}\cr}\eqno(15.6)$$
Therefore, we may write 
$$\eqalign{p_A^{i,\Out}\,&=\,\l C\low{AB}x_B^{i,\In}\,;\cr
p_A^{i,\In}\,&=\,-\l C\low{AB}x_B^{i,\Out}\,,\cr}\eqno(15.7)$$
where summation over $B$ is implied, and
$$\eqalignno{[p_A^{i,\In},\,p_B^{j,\Out}]\,&=\,-i\l\d^{ij}C_{AB}\,;&(15.8)\cr
[x_A^{i,\In},\,x_B^{j,\Out}]\,&=\,-i\l^{-1}(C^{-1})\low{AB}\,.&(15.9)\cr}$$
The inverse $C^{-1}$ of the matrix $C$ is actually ill-defined because
$C$ has a vanishing eigenvalue (corresponding to the constant
function), but this can be cured as usual with an infrared cut-off; 
for finite black holes the horizon curvature term comes to the rescue.

In the continuum limit, these equations approach to (14.1 -- 10), as
well as (14.20). Now, since ${\bf x}_A^\In-{\bf x}_B^\In$ and ${\bf
x}_A^\Out- {\bf x}_B^\Out$ for two linked points $A$ and $B$ do not
commute, they fluctuate in accordance with an uncertainty relation,
such that the average link length fluctuates around the Planck length.
Thus, we automatically obtain a finite transverse cut-off, as was
promised at the beginning of this section.

Next, consider a black hole for which both $|{\bf x}^\In|$ and
$|{\bf x}^\Out|$ are bounded, by writing
$$\sum_A|{\bf x}^\Inout|^2\,<\,(2Gm\BH)^2\,.\eqno(15.10)$$
Upon diagonalizing $C_{AB}$, we find at each eigenmode a harmonic
oscillator with a bound on its Hamiltonian. Consequently, this system has
a bound on its total number of states. The bound one finds this way, however, 
is not yet stringent enough to yield the Bekenstein-Hawking entropy
formula.

A different procedure was introduced in Ref\ref{ 30}. As our starting point
we again use Eqs.~(14.1) --~(14.5), but assume these to be valid only
when the functions $U_\In(\tl x)$ and $U_\Out(\tl x)$ are slowly
varying.  For later convenience, we rename the transverse coordinates
on the horizon as $(\s^1,\,\s^2)$, and now define a 2-surface
$x^\m(\tl\s)$ embedded in 4-space:  $$ x^+\,=\,U_\In\ ,\qquad
x^-\,=\,U_\Out\ ,\qquad \tl x\,=\,\tl\s\ ,\eqno(15.11)$$
thus, in contradistinction with the previous procedure, we now keep the
the transverse coordinates for the in- and out-states identical.

The orientation of the surface is given by the tensor
$$W^{\m\n}(\tl\s)\,=\,-W^{\n\m}\,=\,\e^{ab}\,{\pa x^\m\over\pa\s^a}\,
{\pa x^\n\over\pa\s^b}\,.\eqno(15.12)$$ We have
$${\pa\tl x^a\over\pa\s^b}\,=\,\d^a_b\,.\eqno(15.13)$$ 
Now first consider the case that  $x^\pm$
are slowly varying. This implies
$$\eqalign{W^{12}\,=\,1\ ;\qquad& W^{1\pm}\,=\,{\pa x^\pm\over\pa\s^2}\ ;\cr
W^{2\pm}\,=\,-{\pa x^\pm\over\pa\s^1}\ ;\qquad& W^{+-}\,=\,\OO(\pa_\s
x^\pm)^2\ .\cr}\eqno(15.14)$$
Commutation rules follow from Eqs.~(14.2) and~(14.5):
$$\eqalign{[W^{1+}(\tl\s),\,W^{2-}(\tl\s')]\,=\,[W^{2+}(\tl\s),&
\,W^{1-}(\tl\s')]\,=\,i\pa_1\pa_{\,2}f(\tl\s-\tl\s')\,;\cr
[W^{1+}(\tl\s),\,W^{1-}(\tl\s')]\,&=\,-i{\pa^{\,2}\over\pa{\s^2}^2}f(\tl\s-\tl\s')
\ ;\cr [W^{2+}(\tl\s),\,W^{2-}(\tl\s')]\,&=\,-i{\pa^{\,2}\over\pa{\s^1}^2}
f(\tl\s-\tl\s')\ .\cr}\eqno(15.15)$$

As a special case, we have
$$[W^{\m+}(\tl\s),\,W^{\m-}(\tl\s')]\,=\,-i\tds
f(\tl\s-\tl\s')\,=\,i\,\d^2 (\tl\s-\tl\s')\,,\eqno(15.16)$$ where the
index $\m$ is summed over.  It is this equation that we can reformulate
in a manifestly Lorentz covariant form. One then may hope that not only
the longitudinal, but also the shifts in all other directions  will
have been accommodated for.  Since according to Eq.~(15.14), $W^{12}$
is the dominating component of the tensor $W^{\m\n}$, one may rewrite
the right hand side of Eq.~(15.16) as
$$\e^{+-12}W_{12}(\tl\s)\d^2(\tl\s-\tl\s')\,\approx\,\half\e^{+-\m\n}W_{\m\n}
(\tl\s)\d^2(\tl\s-\tl\s')\,,\qquad\hbox{with}\quad\e^{+-12}\,=\,
i\e^{3412}\,=\,i\,.\eqno(15.17)$$
The covariant generalization is then:
$$[W^{\m\a}(\tl\s),\,W^{\m\b}(\tl\s')]\,=\,\half\d^2(\tl\s-\tl\s')\e
^{\a\b\m\n}W^{\m\n}(\tl\s)\,.\eqno(15.18)$$
This equation, as well as (15.16), is invariant under all continuous
reparametrizations of the $\tl\s$ coordinates (note that $W^{\m\n}$, as
defined by Eq.~(15.12), transforms as a density.)

It is tempting to assume Eq.~(15.18) to have a wider range of validity
than the non-covariant Eqs.~(14.1) -- (14.10). After all, Lorentz
invariance guarantees that Eq.~(15.18) continues to hold when the
derivatives of $x^{\pm}(\tl x)$ are arbitrarily large. Unfortunately,
the equations~(15.16) do not form a closed algebra, since at the left
hand side the index $\m$ is still summed over.  One can, however, limit
oneself to the self-dual part:

$$K^{\m\n}\,=\,i(W^{\m\n}+\half\e^{\m\n\k\l}W^{\k\l})\,,\eqno(15.19)$$
which has only three independent components:
$$K_1\,=\,i(W^{23}+W^{14})\ ;\qquad
K_2\,=\,i(W^{31}+W^{24})\ ;\qquad
K_3\,=\,i(W^{12}+W^{34})\ ,\eqno(15.20)$$
and, indeed, their algebra closes. From Eq.~(15.19) we derive:
$$[K_a(\tl\s),\,K_b(\tl\s')]\,=\,i\e_{abc}K_c(\tl\s)\d^2(\tl\s-\tl\s')\,.
\eqno(15.21)$$

The operators $K_a(\tl\s)$ are distributions. In order to construct
representations of the algebra~(15.21), we introduce test functions
$f(\tl\s)$, $g(\tl\s)$, and write $$\eqalignno{L_a^{(f)}\,&\ddef\,\int
K_a(\tl\s)f(\tl\s)\dd^2\tl\s\,,&(15.22)\cr
[L_a^{(f)},\,&L_b^{(g)}]\,=\,i\e_{abc}L_c^{(fg)}\,.&(15.23)\cr}$$
Restricting oneself to test functions $f$ with $f^2=f$, which only take
the values 0 or 1, we find that the operators $L_a^{(f)}$ obey the
commutation rules of the angular momenta:

$$[L_a^{(f)},\,L_b^{(f)}]\,=\,i\e_{abc}L_c^{(f)}\,.\eqno(15.24)$$ Notice
that, since these operators $L_a$ are obtained by integrating $K_a$
over the region(s) where $f=1$, and because the definition of $K_a$ can
be traced back to Eq.~(15.12), one can rewrite $L_a^{(f)}$ as a contour
integral:  $$L_1^{(f)}\,=\,i\oint\limits_{\d f}(x^2\dd x^3+x^1\dd
x^4)\,,\qquad\hbox{etc.}\,, \eqno(15.25)$$
where $\d f$ stands for the boundary of the support of $f$.

\midinsert\epsffile{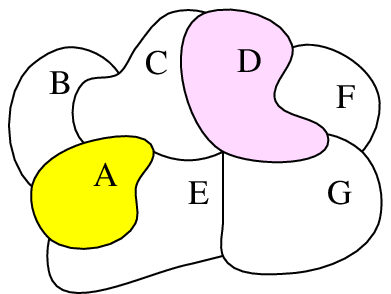}\scrunch\narrower\cl{Figure 7. Domains
on the horizon corresponding to a representation of the algebra
(15.21).}\endinsert

Suppose now that we have a set of test functions $f$ which are equal to
1 on domains $A$ or $B$, etc., and zero elsewhere. The domains form a
lattice (of our choice) on the horizon, see Fig.~7. In each domain we
have a set of three operators $L_a$ that commute as angular momentum
operators. The states could be formed out of the $|\ell,m\ket$
eigenstates of ${\bf L}^2$ and $L_3$. If we combine domains to form
some larger domain, the corresponding angular momentum operators must
be added to form the new ${\bf L}$ operators, by the use of
Clebsh-Gordan coefficients.  Actually, if any of the $\ell$ values is
larger than the minimal value $\half$ (or perhaps, in some cases, 1),
one can imagine splitting the corresponding domain into smaller ones
with each the $\ell$ value $\half$. Thus, one may end up with a lattice
where on each site one has $m=\pm\half$. It would not make much sense
to maintain domains which have ${\bf L}=0$, because the vanishing of
the integrals (15.25) would imply that these regions have no spacial
extent.

At first sight, this looks like a complete resolution of our problems.
If each domain could be attributed an area equal to $4G\ln 2$ (see
Eq.~(7.12)), we exactly reproduce Eq.~(7.11) for the level density.
Unfortunately, life is not so simple. In Eq.~(15.20), $W^{i4}=iW^{i0}$
are antihermitean operators, but $W^{12}$, $W^{23}$ and $W^{31}$ are
hermitean. Therefore, the hermitean conjugates of $K_a$, and those of
$L_a$, are the {\it anti\/}self-dual parts of $W^{\m\n}$.  The $L_a$
operators are not hermitean, and therefore the $\ell$ and $m$ quantum
numbers need not be subjected to the usual constraints of being
half-integer, nor to obey the usual inequalities $|m|\le\ell$.

\secbreak
\ni{\bf 16. CONSEQUENCES FOR THEORIES OF SPACE-TIME}\medskip In a
completely satisfactory theory, the consequences of coordinate
reparametrization invariance (or whatever replaces that), should have
been exploited to the very end. It means that a black hole horizon
could be transformed to exist anywhere in space-time. Flat space-time
should locally be indistinguishable from an infinite size black hole.
This provides us with a dilemma. If the transformation from flat
space-time to Rindler space is one-to-one in terms of the elements of
Hilbert space, and if indeed the horizon only has one Boolean degree of
freedom (such as a spin being $\pm\half$), at every surface element of
size $A_0=4G\ln2$, then also the Hilbert space of ordinary particles in
flat space-time should carry only one Boolean degree of freedom for
each surface element $A_0$. This is a big contrast with theories such
as quantum field theories on lattices, in which the spacelike lattice
is three-dimensional, and hence should show at least one Boolean degree
of freedom for each fundamental volume element.

{\it Information content\/} may be a crucial concept in theories for
Planck scale physics.  The fact that information in a given volume
$V$ must be limited to a boundary surface, can also be deduced as
follows. Let us ask what is the {\it maximal\/} amount of information
that can be stored in {\it any} volume $V=R^3$, given only one constraint:
the total energy in use should be such that, if gravitational collapse
takes place, the resulting black hole should not be larger than $V$,
otherwise the information cannot be retrieved.

First, let us fill $V$ with non-interacting particles. the number of
different quantum states can easily be derived using thermodynamics.
Since $$E\,=\,C_1VT^4\,\lap\,R\ ,\qquad\hbox{and
}\ S\,=\,C_2VT^3\,,\eqno(16.1)$$ where $C_i$ are constants of order one
in natural units, we find $$T\,\lap\,C_3R^{-\half}\ ;\qquad
S\,\lap\,C_4R^\fract32\,\approx C_4\sqrt V\,, \eqno(16.2)$$ The number
of states is then approximately $\exp\, C_4\sqrt V$.

Next, suppose that, due to gravitational attraction, several black
holes were formed, having energies $E_i$. Then
$$E\,=\,\sum_iE_i\,\lap\,R\ ;\qquad S\,=\,\sum_i\pi
R_i^2\,=\,4\pi\sum_iE_i^2\,.  \eqno(16.3)$$ This is bounded by
$$S\,\lap\,4\pi E^2\,\approx\,\pi R^2\,,\eqno(16.4)$$ which equals
$\fract14$ times the surrounding surface. The bound is indeed saturated
if just one black hole is formed that just barely fits inside $V$.
More information inside $V$ is impossible.

The information content of 3-space according to the arguments just
sketched may be compared to that of a holographic
registration\ref{31, 32} of a three dimensional object; the photographic
plate has a resolution not better than the Planck length, showing only
one pixtel that can be either black or white, at every unit
$A_0=4G\ln2$ of its surface. This rendering of our three dimensional
world can be blurry at best, but because the Planck length is extremely
small, nothing of this phenomenon is noticed in ordinary physics.

Attempts to construct quantum field theoretical systems with their
information content limited to any given flat (infinite) surface, were
made\ref{31}, but not with much success. 

\midinsert\epsffile{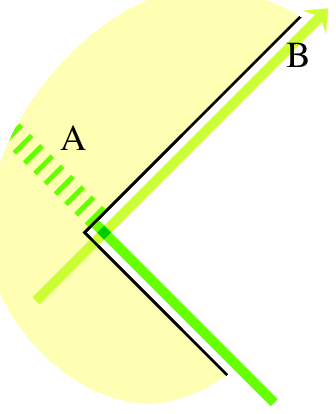}\scrunch\cl{Figure~8. The
ingoing observer ({\smallit A\/}), and the Hawking observer ({\smallit
B\/}).}  \endinsert

Let us return to the argument at the end of Section~9 concerning the notion of
causality\ref{25}.  It has often been raised as a point of criticism against
our scattering matrix Ansatz. See Fig.~8. An observer $A$ passes
through an horizon, while also an onserver $B$ detects Hawking
radiation. If this were flat space-time, these two observers were
considered to be spacelike separated, and therefore their measurement
operators commute. Hilbert space can be factored into a space of states
whose properties can be detected by $A$, and another space of states
whose properties can be detected by $B$, and possible further factors
that can be seen neither by $A$ nor by $B$. If, however, this space
were considered to be the horizon of a black hole, the states seen by
$A$ are related to the states seen by $B$ through an $S$-matrix, and
hence no longer independent. For the black hole physicist, there is no
contradiction.  Any measurement made by $B$, implies the introduction
of states obtained from the Hartle-Hawking vacuum by acting on it with
operators that create or remove particles seen by $B$, which for $A$
would by outrageously energetic.  These particles would cause
gravitational shifts that seriously affect the ingoing objects,
including the fragile detectors used by $A$. Thus, these observations
cannot be independent. What is new here, even for any possible flat
space-time observer, is that trans-Planckian particles are involved
(with this term we mean particles whose energies are far beyond the
Planck value)

Apparently, new phenomena strongly affect the conventional form of
quantum mechanical Hilbert space when trans-Planckian particles enter
the scene. With trans-Planckian particles around, spacelike separated
operators may no longer commute with each other.
 
\secbreak
\ni{\bf 17. OUTLOOK}\medskip
Even though the philosophy, adhered to in this paper, is completely
straightforward, and should not present fundamental conceptual
problems, it nevertheless turned out to be extremely difficult to
implement it completely.  We have not yet been able to fully exploit
the presence of fermions in the Standard Model, which should probably
lead to anticommuting operators on the horizon. The effects of
transverse gravitational shifts were also hard to implement, since
these shifts do not commute with the longitudinal ones (because of
their $\tl x$-dependence). We have not mentioned a third difficulty:
the mass shell conditions for the in- and outgoing particles. We took
these to be essentially massless, but most particles (such as the
electrically charged ones) have a lower bound for their masses.
Transverse momenta and masses, however, cause outgoing particles to fall
back in again. The difficulty connected to this is the fact that, close
to the horizon, ingoing and outgoing states will be difficult to
distinguish. Presumably, the splitting of $S$ according to $S=S\out
S\low{\rm hor}S\in$ (Eq.~11.1), must be further refined.

The resemblance to string theory in our final results may suggest that
one should readdress the black hole using string theory. Some caution
however is called for. It is well-known that string theory requires a
10 or 26~dimensional target space, if tachyons and other unphysical
features are to be avoided, but such arguments do not directly apply to
our present aproach: unitarity and causality look very different, as is
manifest from the observation that our string constant is purely
imaginary.  Secondly, by considering the ``information content" of the
states in our Hilbert space, we infer that a cut-off at the Planck
scale is required that turns our world into a discrete one at that
scale. This is quite unlike the starting points of string theory.
Convergence of the various approaches may well be envisioned, but it is
conceivable that two-dimensional conformal quantum field theory is no
more (or less) relevant here than it is in certain statistical models
such as the Ising model. We should keep in mind that QCD is also a
theory that shows stringlike behavior, but clearly lives in four
space-time dimensions, so that apparently the formal unitarity
arguments are not applicable here.

the observation of black hole--white hole complementarity (Section~11)
suggests an interesting relationship between the {\it black hole
horizon\/} and the {\it white hole singularity\/}, and vice versa.
After all, a white hole singularity would develop as soon as one
allows Hawking particles to produce a gravitational field, as one 
would be tempted to do when contemplating time reversal invariance.
Indeed, the point $S$ in Fig.~2b, is not truly a point, but gets
the extended shape of a caustic when ingoing matter is deprived of its
spherical symmetry. The operator $U_\Out(\tl x)$ could be regarded
as the one describing this caustic. When ingoing matter is allowed
to enter during sufficiently large time intervals, this caustic becomes
a true fractal. At the same time, this fractal may be relavant for
the description of the singularity in the time-reversed black hole.
A duality relationship between the black hole singularity and the
horizon has been proposed in the framework of string theory. 

Numerous other attempts have been made to ``tame" the quantum black
hole.  A possible description of ``quantum hair" can be imagined by
introducing discrete but local symmetries, such as what happens when a
Higgs field with large isospin is introduced in an $SO(3)$ Yang-Mills
theory\ref{33}. Introducing such discrete local symmetries appears to be
a large step, not corroborated by any observations within the frame
of the Standard Model, and in our scheme unneccessary.

Several authors suggest\ref{34} that the black hole horizon area is
quantized into multiples of a Planck-sized unit, perhaps $4G\ln2$.
Now, since the black hole mass carries an imaginary part (see
Section~6), the area is actually not very precisely defined, so that a
quantization rule of this kind may have relatively little physical
impact; by itself, it does not help our understanding very much.
However, when combined with loop quantization of gravity, as advocated
by Ashtekar, Rovelli, Smolin, and others\ref{35}, perhaps new avenues may
be opened.  It is obvious that more new ideas are needed.
\secbreak

\ni{\bf REFERENCES}
 \def\br{\hfil\break}
\def\id{{\it id.}}

\item{(1)} S. Chandrasekhar, {\it The Mathematical  Theory  of
     Black Holes}, Clarendon Press, Oxford University Press.
 
\item{(2)}K.S. Thorne, R.H.~Price and D.A.~MacDonald, {\it Black Holes: the Membrane 
	Paradigm}, Yale Univ. press, New  Haven, 1986.

\item{(3)} S.W. Hawking, Commun. Math. Phys. {\bf 43} (1975) 199; J.B.
Hartle and S.W. Hawking, Phys. Rev. {\bf D13} (1976) 2188.
	 
\item{(4)} S.W. Hawking, Phys. Rev. {\bf D14} (1976) 2460; {\it id.}
 Commun. Math. Phys. {\bf 87} (1982) 395; S.W. Hawking and R. Laflamme,
Phys. Lett. {\bf B209} (1988) 39; \br
D.N. Page, Phys. Rev. Lett. {\bf 44} (1980) 301; \br
R. Haag, H.Narnhofer and U.Stein, Commun. Math. Phys. {\bf 94} (1984) 219; \br
R. Sorkin, Phys. Rev. Lett. 56 (1986) 1885, Phys. Rev. D {\bf 34} (1986) 373; \br
P.~Mitra, {\it Black hole Entropy}, Invited talk delivered at XVIII IAGRG Conference,
Madras, February 1966, hep-th/9603184.

\item{(5)} J.D. Bekenstein, Nuovo Cim. Lett. {\bf 4} (1972) 737;  Phys. Rev. D
{\bf 7} (1973) 2333; D {\bf 9} (1974) 3292.

\item{(6)} R.M. Wald, Commun. Math. Phys. {\bf 45} (1975); \id, Phys.~Rev. 
	D {\bf 20} (1979) 1271; 
W.G. Unruh and R.M. Wald, Phys. Rev. D {\bf 29} (1984) 1047; W.G. Unruh, 
Phys.~Rev. D {\bf 14} (1976) 870; \br
S.W. Hawking and G. Gibbons, Phys. Rev. D {\bf 15} (1977) 2738.

\item{(7)} C. Callan, S. Giddings, J. Harvey and A. Strominger, Phys.
Rev. D {\bf 45} (1992) 1005.

\item{(8)} G. 't Hooft, J. of Geometry and Physics {\bf 1} (1984) 45-52,
\id, in:
{\it Niels Bohr:  Physics and the World}, Proc. of the Niels Bohr Centennial
Symposium, Boston, MA, USA, Nov. 12-14, 1985. Eds. H. Feshbach, T.
Matsui and A Oleson. Publ. by Harwood Academic Publ. GmbH (1988) p.171.

\item{(9)} G. 't Hooft,  Nucl. Phys. {\bf B256} (1985) 727-745. 

\item{(10)} G. 't Hooft,  Physica Scripta, {\bf  T15} (1987) 143-150.

\item{(11)} E.T. Newman et al, J. Math. Phys. {\bf 6} (1965) 918; B. Carter, 
Phys. Rev. {\bf 174} (1968) 1559.
	      
\item{(12)} S.W.  Hawking  and  G.F.R.  Ellis,  {\it The Large Scale
Structure of Space-time}, Cambridge: Cambridge Univ. Press, 1973.

\item{(13)} S.W.~Hawking, G.T.~Horowitz and S.F.~Ross,  Phys. Rev. D 
{\bf 51}, 4302 (1995); \br
C. Teitelboim, Phys. Rev. D {\bf 51} 4315 (1995).
		
\item{(14)} L. Susskind, ``Some speculations about black hole entropy
in string theory", hep-th/9309145; \br G.T.~Horowitz, ``The origin
of black hole entropy in string theory", to appear in the proceedings
of the Pacific Conference on Gravitation and Cosmology, Seoul, Korea,
February~1-6, 1996, gr-qc/9604051; \br J.G.~Russo, ``The string
spectrum on the horizon of a non-extremal black hole", hep-th/9606031.

\item{(15)} J.D. Bekenstein, Nuovo Cim. Lett. {\bf 4} (1972) 737; \id,
Phys.~Rev. D {\bf 7} (1973) 2333, D {\bf 9} (1974) 3292.
	  
\item{(16)} S.~Carlip, ``The statistical mechanics of horizons and
black hole thermodynamics", talk given at the Pacific Conference on
Gravitation and Cosmology, Seoul 1996, gr-qc/9603049; {\it id.}, ``The
statistical mechanics of the three-dimensional euclidean black hole",
gr-qc/9606043.

\item{(17)} W. Rindler, Am.J. Phys. {\bf 34} (1966) 1174.

\item{(18)} P.C. Vaidya, Proc. Ind. Acad. Sci. {\bf A 33} 264 (1951).

\item{(19)} T. Dray and G. 't Hooft, Comm. Math. Phys. {\bf 99} (1985) 613.
 
\item{(20)} P.Bizo\'n, ``How to make a tiny black hole?", Cracow (Poland)
preprint, june 1996 (gr-qc/9606060).

\item{(21)} G. 't Hooft, Acta Physica Polonica {\bf  B19} (1988) 187-202.

\item{(22)} J.-G.~Demers and C.~Kiefer, {\it Decoherence of Black holes by Hawking Radiation},
McGill prepr. 95-56 / Freiburg prepr. THEP-95/22.

\item{(23)} T. Banks, L. Susskind and M.E.~Peskin, Nucl. Phys. {\bf B244} (1984) 125.

\item{(24)} J.D. Bekenstein, Phys. Rev. D {\bf 5} (1972) 1239, 2403.
	 
\item{(25)} C.R. Stephens, G. 't Hooft and B.F. Whiting,  Class. Quantum
Grav. {\bf  11} (1994) 621.

\item{(26)}P.C. Aichelburg and R.U. Sexl, Gen.Rel. and Gravitation {\bf 2} (1971) 303;    
W.B. Bonner, Commun. Math. Phys. {\bf 13} (1969) 163.

\item{(27)}  T. Dray and G.~'t~Hooft, Nucl. Phys. {\bf  B253} (1985) 173.

\item{(28)} G. 't Hooft, Proc. of the 4th
seminar on Quantum Gravity, May 25-29, 1987, Moscow, USSR. Eds. M.A.
Markov, V.A.  Berezin and V.P. Frolov, World Scientific 1988, p. 551;
 {\it id.}, ``Black holes as clues to the problem of quantizing 
gravity", Lecture notes for the CCAST/WL Meeting on {\it fields, strings and quantum
gravity}, Beijing, June 1989. (Gordon and Breach, London); \br
 {\it id.}  Nucl. Phys.  {\bf B335} (1990) 138; Physica
Scripta T {\bf 36} (1991) 247. \br
{\it id.}  ``Scattering matrix for a quantized black hole", In book:
{\it Black Hole Physics},  V.~De Sabbata and Z.~Zhang (eds.). 1992 Kluwer
Academic Publishers, The Netherlands, p. 381; 
{\it id.}  ``S-Matrix theory for black holes",  In Proceedings of a NATO
Advanced Study Institute on {\it New Symmetry Principles in Quantum Field
Theory}, held July 16-27, 1991 in Carg\`ese, France. Eds. J. Fr\"ohlich,
G.~`t~Hooft, A.~Jaffe, G.~Mack, P.K.~Mitter and R.~Stora NATO ASI
Series, 1992 Plenum Press, New York, p. 275;
{\it id.}  ``Black Holes, Hawking Radiation and the
Information Paradox", Proceedings of {\it Trends in Astroparticle Physics},
Stockholm, Sweden, Sept. 22-25, 1994, THU-94/20.
 
\item{(29)} N. Itzhaki, ``Some remarks on 't Hooft's $S$-matrix for black holes",
hep-th/9603067.

\item{(30)} G. 't Hooft, ``Black Hole Quantization and a connection to String Theory", in:
{\it Physics, Geometry and Topology}, Banff Summer School in Theoretical
Physics, NATO ASI Series, Ed. H.C. Lee (1990 Plenum Press, New York) p. 105; 
{\it id.} ``More on the black hole $S$-matrix", in: Proceedings of the
Fifth Seminar {\it Quantum Gravity}, Moscow, USSR, 28 May-1 June 1990.
M.A. Markov, V.A. Berezin and V.P.~Frolov (eds.). World Scientific,
Singapore (1991) p. 251.

\item{(31)} G.~'t~Hooft, ``Dimensional Reduction in Quantum Gravity",
in {\it Salamfestschrift: a collection of talks}, World Scientific Series in
20th Century Physics, vol.~4, ed. A.~Ali, J.~Ellis and S.~Randjbar-Daemi 
(World Scientific, 1993), THU-93/26,  gr-qc/9310026.

\item{(32)}   L.~Susskind, L.~Thorlacius and J.~Uglum, Phys. Rev. D {\bf 48} (1993) 3743
(hep-th/9306069); L.~Susskind, ``The world as a hologram", J. Math. Phys. {\bf 36} (1995) 6377, hep-th/9409089;\br
S. Corley and T. Jacobson, ``Focusing and the holographic hypothesis", gr-qc/9602031.

\item{(33)} L.M.~Krauss and F.~Wilczek, Phys. Rev. Lett. {\bf 62} (1989) 1221; J.~Preskill and 
L.M.~Krauss, Nucl. Phys. {\bf B341} (1990) 50; L.M.~Krauss, Gen.
Rel. Grav. {\bf 22} (1990). \br
A.~Strominger, ``Statistical Hair on Black Holes", Rutgers prepr. RU-96-47,
hep-th/9606016.

\item{(34)} J.D.~Bekenstein and V.F.~Mukhanov, Phys. Lett. B {\bf 360} (1995) 7.

\item{(35)} A. Ashtekar, Phys. Rev. D {\bf 36} (1987)  1587;  A.  Ashtekar
et  al,  Class.
     Quantum Grav. {\bf 6} (1989) L185; \br C. Rovelli, INFN  Roma  preprint,
     May 1988; C. Rovelli and L. Smolin, INFN Roma preprint  N.  620,
     Sept.  1988.

\item{(36)} L.P. Kadanoff and H. Ceva, Phys. Rev. B {\bf 3} (1971) 3918.

\item{(37)} G.~'t~Hooft, Acta Physica Austr., Suppl. XXII (1980) 531; \id, Nucl. Phys. 
{\bf B138} (1978) 1.                                 
\bye